\documentclass[letterpaper,english,aps,pra,twocolumn,showpacs,superscriptaddress]{revtex4}

\usepackage[T1]{fontenc} 
\usepackage[latin9]{inputenc}
\usepackage{color} 
\usepackage{verbatim} 
\usepackage{amsmath}
\usepackage{graphicx} 
\usepackage{amssymb}

\makeatletter

\@ifundefined{textcolor}{}
{%
 \definecolor{BLACK}{gray}{0}
 \definecolor{WHITE}{gray}{1}
 \definecolor{RED}{rgb}{1,0,0}
 \definecolor{GREEN}{rgb}{0,1,0}
 \definecolor{BLUE}{rgb}{0,0,1}
 \definecolor{CYAN}{cmyk}{1,0,0,0}
 \definecolor{MAGENTA}{cmyk}{0,1,0,0}
 \definecolor{YELLOW}{cmyk}{0,0,1,0} }

\makeatletter

\usepackage{ae}
\usepackage{aecompl}
\usepackage{subfigure}
\usepackage{bm}

\@ifundefined{definecolor}
 {\@ifundefined{definecolor}
 {\usepackage{color}}{}  }{}

\newcommand{\ket}[1]{\left\vert#1\right\rangle}
\newcommand{\bra}[1]{\left\langle#1\right\vert}

\makeatother
\usepackage{babel}

\begin{document}

\title{Pointer States via Engineered Dissipation}

\author{Kaveh Khodjasteh} 
\affiliation{\mbox{Department of Physics and Astronomy, 
Dartmouth College, Hanover, NH 03755, USA}}

\author{Viatcheslav V. Dobrovitski} \affiliation{\mbox{Ames
Laboratory, Iowa State University, Ames, IA 50011, USA}}

\author{Lorenza Viola}
\affiliation{\mbox{Department of Physics and Astronomy, 
Dartmouth College, Hanover, NH 03755, USA}}

\begin{abstract}
Pointer states are long-lasting high-fidelity states in open quantum
systems. We show how any pure state in a
non-Markovian open quantum system can be made to behave as a pointer
state by suitably engineering the coupling to the environment via
open-loop periodic control.  Engineered pointer states are constructed
as {\em approximate fixed points} of the controlled open-system
dynamics, in such a way that they are {\em guaranteed} to survive over
a long time with a fidelity determined by the relative precision with
which the dynamics is engineered.  We provide quantitative
minimum-fidelity bounds by identifying symmetry and ergodicity
conditions that the decoherence-inducing perturbation must obey in the
presence of control, and develop explicit pulse sequences for
engineering any desired set of orthogonal states as pointer states.
These general control protocols are validated through exact numerical
simulations as well as semi-classical approximations in realistic
single- and two- qubit dissipative systems.  We also examine the role
of control imperfections, and show that while pointer-state
engineering protocols are highly robust in the presence of systematic
pulse errors, the latter can also lead to unintended pointer-state
generation in dynamical decoupling implementations, explaining the
initial-state selectivity observed in recent experiments.
\end{abstract}


\date{\today}
\maketitle

\section{Introduction}
\label{sec:Intro}

Understanding how and to what extent robust classical properties can
dynamically emerge in open quantum systems as a result of the
unavoidable interaction with the surrounding environment has long been
identified as a problem of central significance to quantum physics and
quantum engineering.  The concept of {\em pointer states} (PSs) has
been introduced by Zurek \cite{Zurek81} to capture the fact that not
all initial pure states of an open quantum system may be equally
fragile with respect to the interaction with the environment: PSs are
distinguished by their ability to persist with high fidelity over time
scales of practical interest, and are thus natural candidates to
describe ``preferred'' states in which open quantum systems are found
in reality.  As a result, PSs play a fundamental role in
investigations of the quantum-to-classical transition and quantum
measurement models \cite{RevModPhys.75.715}, as well as of general
aspects of ``most classical'' minimum-uncertainty states in
quantum-dynamical systems \cite{Habib93,Isar,Eisert2004PS,Sergio}.  In
the context of quantum information processing, a set of mutually
orthogonal PSs (a {\em pointer basis}) provides the simplest example
of an ``information-preserving structure'' (IPS) \cite{IPS}: since
arbitrary convex mixtures of PSs are preserved, a pointer basis
naturally realizes a robust \emph{classical memory}.  As a result, PSs
are also practically attractive in view of their potential for
long-lasting storage capabilities.

From a physical standpoint, the robustness of PSs can be traced back
to the fact that they become ``least entangled'' with the environment
in the course of the dynamics \cite{Zurek81}.  Mathematically (in a
sense that will be made more precise later), this is only possible if
the open-system Hamiltonian exhibits a sufficient degree of {\em
symmetry}, which effectively allows PSs to be eigenstates (fixed
points \cite{IPS}) in the resulting system-plus-bath dilation.  This
has two implications: On the one hand, for a {\em generic} open
quantum system, such a symmetry is {\em not} typical and at best
approximate, thus PSs need not exist -- with all the initial
preparations of the system being rapidly degraded over comparable time
scales.  On the other hand, even in situations where a robust set of
states would be naturally ``ein-selected'' over time, the latter would
be inflexibly determined by the Hamiltonian under consideration --
with the resulting PSs not necessarily coinciding with states of
interest, and with no control over their actual fidelity and lifetime.
By reversing this logic, we may then ask whether by suitably
``engineering dissipation'', that is, by resorting to external
manipulation and perfection of the symmetry in the controlled
open-system Hamiltonian, it is possible to make {\em any target set of
initial pure states into artificial PSs}, so that the desired initial
preparations can be robustly stored over time and their features
retrieved on demand.  This is the question that will be addressed and
constructively answered in this paper -- a task to which we refer to
as ``pointer state engineering''.

Our strategy relies on open-loop (feedback-free) quantum control
methods, close in spirit to dynamical decoupling (DD) approaches to
decoherence suppression
\cite{Viola1998,Viola1999Dec,ZanardiSym0,wocjan2001universal,Viola2002,Khodjasteh2004,Kurizki,Uhrig2007,WestFongLidar2010}
and robust quantum computation
\cite{Viola1999Control,Viola2003Euler,Khodjasteh-Hybrid,Khodjasteh2010dcg}.
The idea is to start from the ``bare'' open-system Hamiltonian that
describes the interaction between the system and its environment, and
to incorporate this Hamiltonian along into a pre-designated control
recipe that acts directly {\em only} on the system, either in the form
of sufficiently fast sequences of control pulses or continuous
time-dependent modulation.  Since in practice only limited knowledge
of the bath degrees of freedom may be available, an important
requirement in DD constructions is that they be robust against
variations in the bath operators, and determined only by algebraic
properties of the underlying open-system Hamiltonian.  While a variety
of different protocols exist (and are being successfully tested in the
laboratory \cite{DDexp,Tyryshkin2010}), their common aim is the
synthesis, perfection, and upkeep of an effective Hamiltonian which is
``dynamically symmetrized'', so that interactions with the environment
are \emph{removed}, up to a given order of accuracy, for the evolution
of any state inside the entire system's Hilbert space
\cite{ZanardiSym0,Viola1999Dec} or within a (control-dependent)
``dynamically generated'' decoherence-free subspace or noiseless
subsystem \cite{Viola2000Dygen,Wu-Create}.  Technically, this is
achieved via a number of algebraic techniques for manipulating generic
interaction Hamiltonians, along with analytic approaches for
perturbatively (or numerically \cite{Biercuk09,Uys09,badd2010})
reducing the unwanted decoherence contributions and bounding the
residual errors.

While many of the building blocks used in the design and analysis of
DD protocols will also be employed for the task of engineering PSs, it
is important to clarify how the two problems differ.  In a typical DD
setting, the goal is to synthesize, for a given system-bath
Hamiltonian, a target unitary propagator (the identity evolution) on
the system with a sufficiently high gate fidelity, so that, ideally,
{\em arbitrary initial preparations} can be robustly preserved over a
desired storage time (or stroboscopically in time if the DD cycle is
repeated).  As a consequence, in a good DD scheme the control
performance should be as {\em unbiased} as possible with respect to
different initializations, and DD protocols must be constructed
without making reference to and/or taking advantage of possible
knowledge of the system's initial state.  While the system-bath
Hamiltonian is also given in the PS-engineering problem, the target
set of initial preparations to be preserved is also specified as an
additional input.  Thus, knowledge of this target set should be
explicitly incorporated into designing a good PS-protocol, so that the
output fidelity is {\em optimized} for the desired PSs over the
desired storage time. These differences result in important practical
advantages: while DD methods discovered so far can in fact only
guarantee a reduced fidelity \emph{decay rate} over time, in
PS-engineering we will be able to guarantee a high fidelity
\emph{value }over long time-spans. The end result is that the
designated states can live much longer (in principle as long as
desired) in a PS-engineering scheme rather than in general DD
procedures intended for quantum memory.

A few remarks may be useful to further place our work in context.
First, we note that, interestingly, high-order/uncanceled corrections
to effective Hamiltonians in DD constructions may contain an
additional degree of symmetry (with respect to the minimum needed to
ensure the intended averaging of the system-bath interaction) and thus
result in the generation of ``accidental'' PSs -- and effective {\em
decoherence freezing}, as predicted in \cite{Wen01,Wen08}.
Remarkably, such accidental PSs have been observed in recent DD
experiments \cite{Tyryshkin2010,Wang2010}.  While the physical
mechanism responsible for the observed initial-state-sensitivity is
different than in a PS-engineering protocol (stemming, in the
experiment, from systematic pulse errors), we shall also address the
emergence of such accidental PSs in DD sequences and show how they may
be formally related within a unified control-theoretic framework.
Second, we also iterate that DD has been invoked as a general strategy
for generating IPSs in (non-Markovian) open quantum systems --
decoherence-free subspaces and noiseless subsystem
\cite{Viola2000Dygen,Wu-Create}, which can both be seen as
multi-dimensional generalizations of PSs in an appropriate sense
\cite{IPS}.  Again, a key difference is that such schemes are not
tailored to preserve a target subspace or subsystem specified in
advance. In this sense, our present analysis shares some motivation
with (closed-loop) stabilization protocols for Markovian evolutions
\cite{Wang2001,TicozziTAC,TicozziAutomatica}, restricted however to
purely unitary control resources and to the (simplest) case of {\em
discrete} sets of states ({\em classical} IPS) in the system Hilbert
space.  Lastly, we recall that the possibility of a prolonged
high-fidelity regime -- so-called {\em quantum fidelity freeze} -- has
been extensively analyzed in the context of Loschmidt echoes for
closed quantum systems evolving under a {\em vanishing} time-averaged
perturbation \cite{prosen2003,prosen2005,Weinstein-Eigen} and, in
turn, shown to be intimately related to DD \cite{prosenRev}.  From
this point of view, PS engineering may be interpreted as dynamically
providing the requirements for a stronger freeze phenomenon to emerge,
in generic open quantum systems, for the target set of initial
preparations of interest.

The content of this paper is organized as follows. In
Sec. \ref{sec:Setup}, we describe the general control setting, along
with the algebraic conditions required for the effective controlled
Hamiltonian to admit PSs. In Sec. \ref{sec:Ergodic-Averaging-Induced}
we characterize the effect that the unavoidable deviation of the
actual effective Hamiltonian from the intended PS-supporting form
implies on the quality of the engineered PSs.  Our main result is a
quantitative lower bound for the minimum fidelity which can be
guaranteed for the PSs over a range of storage times in terms of both
spectral/locality properties of the open-system Hamiltonian and
details of the applied control scheme.  Constructive protocols for
engineering an arbitrary set of PSs are described in
Sec. \ref{sec:Control-Sequences}, whereas explicit illustrations in
paradigmatic control scenarios are given in Sec. \ref{sec:Examples}
based on both semiclassical analytical results and exact numerical
simulations.  In particular, single-qubit sequences are analyzed and
contrasted to DD sequences resulting in accidental PSs, and sequences
for engineering Bell states as PSs in two exchange-coupled qubits are
presented, recovering and extending partial results in
\cite{Mukhtar2010}.  Sec. \ref{sec:imperfect} addresses the impact of
limited control as resulting from imperfect initialization
capabilities and/or imperfect control operations.  We both
characterize the degree of robustness of the PS-engineering sequences
against systematic control errors, and, conversely, show how such
control errors can result in accidental PSs in realistic DD sequences.
Sec. \ref{sec:Conclusion} concludes with a summary of our main
results, along with a discussion of their possible implications and
further open problems.

\section{Notation and Problem Setup}
\label{sec:Setup}

Throughout this work, we focus on a finite-dimensional open quantum
system, that is, a distinguished subsystem $S$ with associated
$D_{S}$-dimensional Hilbert space $\mathcal{H}_{S}$, coupled to an
environment (or bath) $B$, with associated $D_{B}$-dimensional Hilbert
space $\mathcal{H}_{B}$ \cite{infinitedim}.  We use $\ket{j}$ to refer
to system states and $\ket{b}$ to refer to bath states, the overall
composite Hilbert space being given by
$\mathcal{H}_{S}\otimes\mathcal{H}_{B}$.  For a product state
$\ket{j}\otimes\ket{b}$, we refer to $\ket{j}$ ($\ket{b})$ as the
system (bath) component, respectively.  A factorizable density
operator will have system and bath components in a similar way.
Unless otherwise stated, we choose units in which $\hbar=1$.  We also
tend to drop trivial tensor product signs, identity operators on $S$
($I_{S}$) or $B$ ($I_{B}$), and ignore ordering of commuting operators
as long as there is no ambiguity.  Finally, we use the asymptotic
notation with respect to operators in a lax manner. For example,
$A+O(\epsilon)$ might refer to an operator $A$ plus corrections whose
norm (such as the maximum singular value norm) is $O(\epsilon)$. In
perturbation theory, $O(x)$ will symbolically denote corrections that
scale linearly with a quantity $x$ which need not be dimensionless.

The joint evolution of the system and the environment is taken to be
generated by a Hamiltonian of the form 
\begin{eqnarray*}
H(t)= H_0 + H_{\text{ctrl}}(t).
\end{eqnarray*}
Here, $H_{\text{ctrl}}(t)$ is a controllable Hamiltonian that acts
non-trivially only on $\mathcal{H}_{S}$ and that shall be employed for
generating unitary operations, whereas the free Hamiltonian $H_0$ is
specified in a frame where no explicit time-dependence is present and
can be further expressed as
\begin{eqnarray}
H_0 & = & H_{SB} +I_{S}\otimes H_B \label{eq:H0} \\ & \equiv &
H_{S}\otimes I_{B} + H'_{SB} + I_{S}\otimes H_B. \nonumber
\end{eqnarray}
That is, $H_{B}$ denotes the internal evolution of the bath, and
$H_{SB}$ denotes the interaction Hamiltonian between $S$ and $B$,
including the internal evolution $H_{S}$ of $S$.  Formally, the latter
can be isolated by demanding that $H'_{SB}$ involves only traceless
operators on $B$.  In this way, the limit of a closed (unitarily
evolving) quantum system is recovered for $H'_{SB}=0$ (thus
$H_{SB}\equiv H_S$).  The limit where the bath is treated
semi-classically corresponds instead to $\mathcal{H}_{B}\simeq
{\mathbb C}$ and the net effect of $H_{SB}$ is a random modification
of the system Hamiltonian, hence formally $H_{SB}$ acts on
$\mathcal{H}_{S}$ only \cite{slowlyf}.

Our goal is to use open-loop control, specifically through the
application of a pre-determined $H_{\text{ctrl}}(t)$, to enhance and
maintain the fidelity of an arbitrary target set of orthogonal PSs.
The strategy we follow is to modify the evolution of the composite
system by subjecting it to {\em repeated identical control cycles}.
Let the evolution propagator for each period (cycle) of the controlled
evolution be denoted by $U_{c}$. Our fundamental assumption is that
the cycle propagator $U_{c}$, no matter how implemented, remains the
same for all cycles applied.  Suppose that the duration of each cycle
is $T_{c}$ and $N$ repetitions are implemented, up to the total time
$T=NT_{c}$, with $N$ integer.  Then the cycle propagator $U_{c}$ {\em
defines} an effective cycle Hamiltonian $H_{c}$ on
$\mathcal{H}_{S}\otimes\mathcal{H}_{B}$ with the following structure:
\begin{eqnarray}
&& U_c(T_c)\equiv U_{c} = \exp[-iT_{c}H_{c}],\nonumber \\ &&
H_{c}=H_{SB,c}+I_{S}\otimes H_{B,c},
\label{eq:ucyc}
\end{eqnarray}
where, in analogy with Eq. (\ref{eq:H0}), $H_{B,c}$ acts as an
internal bath Hamiltonian and $H_{SB,c}$ as a system-bath interaction,
respectively.  We emphasize that $H_{SB,c}$ and $H_{B,c}$ are {\em
not}, in general, the same that appear in the bare interaction in
Eq. (\ref{eq:H0}).  In fact, $H_{SB,c}$ can be made much smaller than
$H_{SB}$ in DD, whereas pure-bath terms may be induced by the control
even for a ``non-dynamical'' bath, for which the bare Hamiltonian
$H_B=0$.  The periodicity of the evolution allows us to obtain a
time-independent effective Hamiltonian $H_{c}$, since stroboscopically
the evolution is given by
\[ U_N(T) \equiv U_{N}=(U_{c})^{N}= \exp[-iTH_{c}].\]
\noindent 
The key to achieve our control objective is to design
$H_{\text{ctrl}}(t)$ in such a way that $H_{SB,c}$ takes a special
form that we define next. The actual control schemes that result in
this special form will be given in Sec. \ref{sec:Control-Sequences}.

Let the orthonormal set of states $\{\ket{j}\}_{j=0}^{p-1}$ ($p\le
D_{S}$) denote our target PSs.  
The basic step is to ensure that $H_{SB,c}$ is expressible in terms of
a {\em dominant} operator $H_{\text{dom}}$ and a {\em perturbation}
$\epsilon H_{\text{per}}$, such that
\begin{eqnarray}
H_{SB,c} (\epsilon) & = & H_{\text{dom}}+\epsilon H_{\text{per}},
\nonumber \\ H_{\text{dom}} & = &
\sum_{j=0}^{p-1}\ket{j}\!\bra{j}\otimes B_{j}+H^{\prime},
\label{eq:eigexp}
\end{eqnarray} 
where $H^{\prime}$ annihilates $\text{Span}\{\ket{j}_{j=0}^{p-1}\}$
and $B_{j}$ are distinct but otherwise arbitrary operators on $B$.
This dependence of bath operators upon $j$ is crucial to ensure that
in the limit where $\epsilon \rightarrow 0$, each $\ket{j}$ (but no
coherent superposition) is {\em invariant} under evolution generated
by $H_c$.  Accordingly, we shall also refer to Eq. (\ref{eq:eigexp})
as the ``PS condition'' henceforth.
For nonvanishing $\epsilon$, the term $\epsilon H_{\text{per}}$ is, in
general, a system-bath Hermitian operator.  Without loss of
generality, we can shift any diagonal contribution of $H_{\text{per}}$
in the PS basis into $H_{\text{dom}}$, so that \cite{residual}
\begin{equation}
\langle j|   H_{\text{per}}|j \rangle  = 0, \quad \forall j.
\label{residual}
\end{equation}
Note that if all but one PSs appear in the sum in
Eq. (\ref{eq:eigexp}), the remaining state will automatically become
part of the sum also, implying the equivalence of $p=D_{S}-1$ and
$p=D_{S}$.  If $p< D_{S}-1$, the system Hilbert space is decomposed
into two orthogonal subspaces, one generated by the PS set
$\{\ket{j}\}_{j=0}^{p-1}$, and the rest.  Beyond Eq.
(\ref{eq:eigexp}) and requiring $\epsilon$ to be sufficiently small,
we need not specify the controlled effective Hamiltonian further.

Clearly, if the perturbation $\epsilon H_{\text{per}}$ was zero, and
the system was initialized in any of the PSs, each of these states
would be indefinitely preserved with maximal fidelity over time, just
as the energy eigenstates in a closed-system setting.  Realistically,
however, even with sophisticated control schemes, the correction
$\epsilon H_{\text{per}}$ cannot in general be avoided. A good control
scheme should ensure that the states $\{\ket{j}\}_{j=0}^{p-1}$ are
still singled out for their high-fidelity evolution, and can thus play
the role of \emph{engineered PSs}.  Prior to providing explicit
control schemes that synthesize effective Hamiltonians close to
$H_{\text{dom}}$, we need to know how the corrections will affect the
survival of the desired PSs in order to be able to maintain their
performance. We thus proceed to analyze the extent to which the fate
of the engineered PSs is modified in the presence of the inevitable
correction terms.

\section{Fidelity Dynamics of Engineered Pointer States}
\label{sec:Ergodic-Averaging-Induced}

As a basic motivating example, consider a two-level system, in which
$H_{\text{dom}}$ is perturbed by a fixed intra-level coupling
$\epsilon H_{\text{per}}$.  If the energy levels are non-degenerate,
the eigenstates of the \emph{unperturbed} Hamiltonian $H_{\text{dom}}$
can be seen as perturbations of the eigenstates of the
\emph{perturbed} Hamiltonian.  Similar to the familiar Rabi problem,
the discrepancy between the eigenstates is controlled by the
\emph{ratio} of the transition term to the energy gap and sets the
amplitude of the oscillations that will ensue.  For a degenerate
system, the unperturbed eigenstates can end up being very far from the
perturbed eigenstates and thus oscillate with a large amplitude
\cite[p. 194]{bhatiaBook}. The eigenstates of the original unperturbed
Hamiltonian $H_{\text{dom}}$ thus play the role of PSs which retain
their high fidelity (modulo some oscillation) under perturbation.

In an open-system setting, we shall use a similar idea with some
modifications.  The eigenstates of a perturbed system-bath Hamiltonian
will still be close to the unperturbed eigenstates as long as there is
no degeneracy, but the actual difference between the eigenstates is
expanded over the composite Hilbert space which can sample over many
basis states. The fidelity distance will thus need to reflect the
larger size of the composite system-bath Hilbert space. This distance
argument can be modified using an ergodic argument to yield a tighter
fidelity bound for the pointer basis elements.

\subsection{Semi-Classical Environment}
\label{sub:Semi-Classical-Bath}

Consider first a semi-classical setting in which, formally,
$H_{B,c}=0$ and $H_{SB,c}$ is a \emph{system} operator. Let
$\{\ket{j(\epsilon)}\}_{k=0}^{D_{S}-1}$ denote the (perturbed)
normalized eigenstates of $H_{SB,c}(\epsilon)=H_{\text{dom}}+\epsilon
H_{\text{per}}$, with eigenvalues $\omega_{j}(\epsilon)$.  Thus,
$\lim_{\epsilon\rightarrow0}\ket{j(\epsilon)}=\ket{j}$ and
$\lim_{\epsilon\rightarrow0}\omega_{j}(\epsilon)=\omega_{j}$, where,
without loss of generality, we may let $\omega_{0}(0)\equiv 0$.

Focus for simplicity on the preservation of a single PS, say
$\ket{0}$.  The system starts at
$\rho_{S}(0)=\ket{0}\negmedspace\bra{0}$ and after $N$ cycles evolves
to $\rho_{S}(N)=U_{c}^{N}\rho_{S}(0)U_{c}^{-N}$.  A convenient metric
for quantifying the distance between $\rho_{S}(0)$ and $\rho_{S}(N)$
is the survival probability (or input-output fidelity
\cite{fidelity}), given by
$$f_{N}\equiv \text{Tr} [\rho_{S}(0) \rho_{S}(N)]=\langle
0\vert\rho_{S}(N)\vert0\rangle,$$
\noindent 
which, since $\rho_{S}(0)$ is pure, is simply related to the
corresponding Uhlmann's fidelity $f^U_N$ via $f_{N}=(f^U_N)^2$. The
initial state $\ket{0}$ will oscillate as a function of $N$ with
frequencies given by $T_{c}\omega_{j}(\epsilon)$. In the absence of
degeneracy (for $\omega_0$), the amplitude of oscillation is
determined by the difference between $\ket{0}$ and $\ket{0(\epsilon)}$
and is thus controlled by $\epsilon$ through the ratio of the
perturbation to the dominant term. More precisely, upon expanding
$\ket{0}$ in the $\{ \ket{k(\epsilon)}\}$ basis and applying $U_c^N$,
the fidelity loss reads as
\begin{eqnarray} 
1-f_{N} & = &
1-\Big\vert\sum_{j}
e^{-i NT_{c}\omega_{j}(\epsilon)} \vert \langle
0|j(\epsilon)\rangle|^2 \Big\vert^{2} .
\label{fid0}
\end{eqnarray}
In the regime where $\epsilon$ is sufficiently small, the required
probability overlap (note that $\vert \langle j|j(\epsilon)\rangle|^2$
is the so-called local density of states \cite{prosenRev}) can be
estimated using standard 1st order non-degenerate perturbation theory,
\begin{eqnarray*}
\left\{
\begin{array}{rcl}
\ket{j(\epsilon)} & = & Z_j^{1/2} \Big( \ket{j} + \epsilon \sum_{n\ne
j} \ket{n} \frac{ \langle n| H_{\text{per}} | j\rangle}{\omega_j
-\omega_n} \Big) , \\ 
Z_j & \approx & 1 - \epsilon^2 \sum_{n\ne j}
\frac{ \vert \langle n| H_{\text{per}} | j\rangle \vert^2}{(\omega_j
-\omega_n)^2}, 
\end{array}\right.
\end{eqnarray*}
yielding
\[ \vert \langle 0 | j(\epsilon)\rangle \vert^2 \approx
\left\{\begin{array}{ll} Z_j, & j = 0 , \\ \epsilon^2 \vert { \langle
0| H_{\text{per}} | j \rangle} \vert^2 / {\omega_j^2}, & j\ne 0.
\end{array} \right. \]
\noindent 
Upon substituting the above expression in Eq. (\ref{fid0}) and 
retaining terms up to order $O(\epsilon^2)$, one finds
\begin{eqnarray*} 
1-f_{N} &\hspace*{-1mm}\approx \hspace*{-1mm} &
\epsilon^2 \sum_{j\ne0} 4 \sin^2 [ {NT_c \omega_j}/{2} ] \frac{\vert
\langle 0| H_{\text{per}}| j\rangle\vert^2}{\omega_j^2}.
\end{eqnarray*}
Under the perturbative assumption that $\vert\langle 0|
H_{\text{per}}| j\rangle/\omega_j\vert$ is sufficiently small for all
$j\ne 0$, this finally results in the desired (uniform) fidelity lower
bound:
\begin{equation} 
f_N \gtrsim 1- 4 \epsilon^{2} D_S \max_{j\ne0} \frac{\vert\langle
0\vert H_{\text{per}}\vert j\rangle\vert ^2}{\omega_j^2}.
\label{eq:semifid}
\end{equation}  
As noted, if a degeneracy exists in $H_{\text{dom}}$, the maximum
fidelity loss cannot be bounded by a smooth function of $\epsilon$,
since $\ket{0(\epsilon)}$ and $\ket{0}$ can be far apart {\em
regardless} of $\epsilon$ and at some point the initial state may
oscillate too far and become completely lost.

In the absence of degeneracy, Eq. (\ref{eq:semifid}) points to an
elementary yet remarkable feature: no matter how long the time passed
since preparation, \emph{the fidelity loss for an approximate
eigenstate is small and bounded}.  This is a crucial property of PSs
and one that we will strive to reproduce in the general quantum
case. As it turns out, the fidelity dynamics of an engineered PS in
the presence of a quantum environment will closely follow the above
perturbative derivation.  Furthermore, the semiclassical approximation
may be of independent interest in realistic settings.  For example,
under appropriate physical and time-scale conditions, the contact
hyperfine interaction of localized electronic spins with the
surrounding nuclear-spin bath in semiconductors can be modeled in
terms of a static but inhomogeneous magnetic field as long as ensemble
measurements are considered
\cite{merkulov2002,Raedt2003,Coish2005,Yao2006,Lucas2009,Wang2010}.
Our semiclassical analysis transcribes perfectly to this setting.

Let us focus, in particular, on a two-level system, and let
$\sigma_{\alpha}$, $\alpha=x,y,z$, denote the corresponding Pauli
matrices.  In the semiclassical limit, let 
\[ H_{SB}= \vec{\mathbf{b}}\cdot \vec{\mathbf{\sigma}} \equiv
b_{x}\sigma_{x}+b_{y}\sigma_{y}+b_{z}\sigma_{z},\]
\noindent 
where $\vec{\mathbf{b}}=(b_{x},b_{y},b_{z})$ is sampled from a
distribution of random vectors. Clearly, $H_{SB}$ itself does not
result in a preserved pointer basis or a preferred direction unless
the distribution of $\mathbf{b}$ is anisotropic. The effective
Hamiltonian, on the other hand, can be engineered as
\begin{equation}
H_{SB,c}=h_{z}(\mathbf{b})\sigma_{z}+\epsilon [
h_{x}(\mathbf{b})\sigma_{x}+  h_{y}(\mathbf{b})\sigma_{y}], 
\label{eq:hsbcstatrand}
\end{equation} 
where $h_{z}(\mathbf{b})\sigma_{z}$ is the dominant term designed to
preserve $\{\ket{0},\ket{1}\}$. The small parameter $\epsilon$ and the
functional forms of $h_{x}(\mathbf{b})$, $h_{y}(\mathbf{b})$, and
$h_{z}(\mathbf{b})$ will depend on the control sequence used for
producing $H_{SB,c}$, as well as the details of the probability
distribution.
The calculation is straightforward for simple control sequences. In
our notation, $\omega_{1}\equiv 2h_{z}(\mathbf{b})$ and
$\vert\langle1\vert H_{\text{per}}\vert0\rangle\vert^{2} \equiv
h_{x}(\mathbf{b})^{2} +h_{y}(\mathbf{b})^{2}$.  Our result for $f_N$
translates into
\begin{eqnarray} 
f_{N} & \hspace*{-1mm}\approx \hspace*{-1mm}&
1-\epsilon^{2}\frac{\left[h_{x}(\mathbf{b})^{2}+h_{y}(\mathbf{b})^{2}
\right]\sin^{2}\left[NT_{c}h_{z}(\mathbf{b})\right]}{4h_{z}(\mathbf{b})^{2}},
\label{eq:fNhxhyhz}
\end{eqnarray}
as long as $h_z( {\bf b}) \ne0$.  Starting from $f_{0}=1$, the
fidelity loss is thus at most
$\epsilon^{2}h_{z}(\mathbf{b})^{-2}[h_{x}(\mathbf{b})^{2}+
h_{y}(\mathbf{b})^{2}]/4$.  Eq. (\ref{eq:fNhxhyhz}) describes the
fidelity dynamics for a single realization of the classical random
field, whereas the actual open-system fidelity is the average of
$f_{N}$ over the distribution of $\mathbf{b}$.  We shall use this
approach in Sec. \ref{sub:Classical-Environment}, to obtain
approximate analytical results to be used in conjunction with our
simulations of decoherence due to a quantum spin bath.  Classical
phase noise will also be relevant to the discussion of DD experiments
on electron spins of P donors in Silicon in Sec. \ref{sub:slava}.

\subsection{Quantum Environment}
\label{sub:Toggling-Frame-of}

In the quantum regime, Eqs. (\ref{eq:ucyc})-(\ref{eq:eigexp}) describe
a bipartite system, evolving under a Hamiltonian of the form
\begin{equation}
H_c= (H_{\text{dom}} + H_{B,c} ) + \epsilon H_{\text{per}} \equiv
H_D + \epsilon H_{\text{per}} .
\label{eq:hd}
\end{equation}
While the dynamics generated by the unperturbed component $H_D$ would
affect a generic initial state preparation, it does {\em not} affect
the fidelity of initial PSs. Consequently, in order to probe the
fidelity dynamics of the PSs only, the basic idea is to isolate the
``PS-preserving dynamics'' by effecting a transformation to a suitable
interaction frame.  As it turns out, this transformation will play a
crucial role in our derivation, as it will provides the basis for
applying von Neumann's mean ergodic theorem (MET) {[}see Appendix
\ref{sec:Mean-Ergodic-Theorem}{]}.

Let $U_{D}\equiv\exp(-iT_{c}H_D)$ denote the propagator for the
pure-bath and dominant interaction dynamics. The single-cycle
propagator can then be factored as follows:
\begin{eqnarray}
U_{c} & \hspace*{-0.6mm}\equiv \hspace*{-0.6mm}& \exp(-i\epsilon
E)U_{D},\label{eq:bchapprox}\\ \epsilon E & \hspace*{-0.6mm}=
\hspace*{-0.6mm}& \epsilon
T_{c}H_{\text{per}}+\sum_{m=1}^{\infty}\frac{b_{m}}{m!}[\epsilon
T_{c}H_{\text{per}},T_{c}H_D]_{m} + \epsilon^{2}E^{[2+]}, \nonumber
\end{eqnarray}
where $E^{[2+]}$ refers to corrections of second and higher order in
$T_{c}H_{\text{per}}$, $b_m$ are the Bernoulli numbers, and
$[A,B]_{m}\equiv[\cdots[A,B],\cdots],B]$, with $B$ appearing $m$ times
\cite{Klarsfeld1989}. Note that due to our assumption in
Eq. (\ref{residual}), $E$ is purely off-diagonal in the PS set, modulo
the higher order $E^{[2+]}$ terms.  The propagator for $N$ cycles thus
reads
\begin{eqnarray*}
U_{N} & = & (U_{c})^{N}=[\exp(-i\epsilon
E)U_{D}]^{N}\equiv\tilde{U}_{N}(U_{D})^{N},
\end{eqnarray*}
where the unitary operator $\tilde{U}_{N}$ represents the propagator
in the toggling frame generated by $(U_{D})^{N}$ and can be
approximated by invoking a Magnus expansion \cite{Iserles2002}:
\begin{eqnarray}
\tilde{U}_{N} & \equiv &\exp(-i\Omega_{N}), \;\;
\Omega_{N}=\epsilon\Omega_{N}^{[1]}+\epsilon^{2}\Omega_{N}^{[2+]}, 
\nonumber \\
\Omega_{N}^{[1]} & = &\sum_{n=0}^{N-1}(U_{D})^{-n}E(U_{D})^{n}, 
\label{eq:omega1}\\
\Vert\Omega_{N}^{[2+]}\Vert & = & O(N^{2}\Vert E\Vert^{2})\approx
O(T^{2}\Vert H_{\text{per}}\Vert^{2}) . \nonumber
\end{eqnarray}

The time-discretization implied by the periodic evolution streamlines
the application of von Neumann's MET: the sum in Eq. (\ref{eq:omega1})
{\em projects $E$ onto the commutant of $U_{D}$} and, as long as
$H_{\text{dom}}$ is non-degenerate, to that of $H_{\text{dom}}$.  To
implement this explicitly, let $\ket{j}\otimes\ket{b}_{j}\equiv
\ket{j,b}$ and $\omega_{j,b}$ denote the eigenstates and eigenvalues
of $H_D$, respectively, and let us expand $E$ in the operator basis
induced by $\ket{j,b}$, that is:
\[ E=\sum_{j_{1},b_{1}}\sum_{j_{2},b_{2}}E_{j_{1},b_{1};j_{2},b_{2}}, \]
\noindent 
where $E_{j_{1},b_{1};j_{2},b_{2}}=\bra{j_{1},b_{1}}E\ket{j_{2},b_{2}}
\ket{j_{1},b_{1}}\!\bra{j_{2},b_{2}}$.  Following Appendix
\ref{sec:Mean-Ergodic-Theorem}, we can partition $\Omega_{N}^{[1]}$ in
Eq. (\ref{eq:omega1}) into diagonal components in the $\ket{j,b}$
basis, namely $\Omega_{N}^{\Vert}$ (projected onto the commutant), and
off-diagonal components in the $\ket{j,b}$ basis, namely
$\Omega_{N}^{\perp}$ (projected outside the commutant). The diagonal
components $\Omega_{N}^{\Vert}$\emph{ }vanish as a consequence of
Eq. (\ref{residual}), since $\Omega_{N}^{[1]}$ is related to
$H_{\text{per}}$ via commutators with $H_{\text{dom}}$ which preserve
the commutant structure \cite{sidenote}.  The off-diagonal components
$\Omega_{N}^{\perp}$, on the other hand, do affect the PS fidelities,
but instead of growing linearly with $N$ (or time $T$), they are kept
bounded by the MET. This is the fundamental feature of PS preservation
that guarantees a non-trivial long-time fidelity behavior. We can
obtain an expression for the effective off-diagonal terms after
averaging under the MET [Eq. (\ref{eq:geomsum}) in Appendix
\ref{sec:Mean-Ergodic-Theorem}]: 
\[
\Omega_{N}^{\perp}=\sum_{j_{1}\ne
j_{2},b_{1},b_{2}}\frac{1-e^{i(N+1)T_{c}(\omega_{j_{1},b_{1}}-
\omega_{j_{2},b_{2}})}}{1-e^{iT_{c}(\omega_{j_{1},b_{1}}-
\omega_{j_{2},b_{2}})}}E_{j_{1},b_{1};j_{2},b_{2}},
\]
where each term $E_{j_{1},b_{1};j_{2},b_{2}}$ is rescaled by a factor
that is controlled by the inverse of the energy difference
$\omega_{j_{1},b_{1}}-\omega_{j_{2},b_{2}}$ and can be bounded
independently of $N$. In summary, we can approximate the total
propagator up to time $T$ as:
\[
U_{N}=\exp(-i\epsilon\Omega_{N}^{\perp})(U_{D})^{N}+
O(\epsilon^{2}T^{2}\Vert H_{\text{per}}\Vert^2),
\]
where 
\[ \Vert\Omega_{N}^{\perp}\Vert\le\frac{\Vert
E\Vert}{\sin(T_{c}\Delta/2)}, \;\;\; \Delta \equiv \min_{j_{1}\ne
j_{2}}\vert\omega_{j_{1},b_{1}}-\omega_{j_{2},b_{2}}\vert. \]
\noindent 
Notice that transitions within the same $\ket{j}$ sector do {\em not}
appear at all, since $E$ has no matrix elements within the same
$\ket{j}$ sector, ultimately due to Eq. (\ref{residual}).

We now proceed to obtain an approximate upper bound for fidelity loss
of the designated PSs, along with conditions for its applicability.
For simplicity, let us as before focus on an initial PS preparation in
$\ket{0}$, that is:
\begin{equation}
\rho(0)=\ket{0}\!\bra{0}\otimes\rho_{B}(0),
\label{eq:rho0}
\end{equation}
where $\rho_{B}(0)$ is an arbitrary state on $B$. The joint state 
after $N$ cycles becomes 
\begin{align*}
\rho(N) & =\exp(-i\epsilon\Omega_{N}^{\perp}) \Big[ \ket{0}\!\bra{0}
\otimes\rho_{B}(N) \Big] \exp(i\epsilon\Omega_{N}^{\perp})\\ &
+O(\epsilon^{2}T^{2}\Vert H_{\text{per}}\Vert^2),
\end{align*}
where $\rho_{B}(N) = U_D^N \rho_B(0) U_D^{-N}$. To focus even more on
the fidelity evolution of the initial state, we can further partition
$\Omega_{N}^{\perp}$ into parts that couple to $\ket{0}$ and the rest:
$\Omega_{N}^{\perp}=\Omega_{N}^{\ket{0}}+\Omega_{N}^{\text{rest}}$.
Up to corrections of
$O(\epsilon^{2}\Vert\Omega_{N}^{\perp}\Vert^{2})$, we can then write
\begin{align}
\rho(N) & =\exp(-i\epsilon\Omega_{N}^{\ket{0}}) \Big[
 \ket{0}\!\bra{0}\otimes\rho_{B}(N) \Big]
 \exp[i\epsilon\Omega_{N}^{\ket{0}})\nonumber \\ &
 +O(\epsilon^{2}T^{2}\Vert H_{\text{per}}\Vert^ 2)\nonumber \\ &
 +O(\epsilon^{2}T_c^2\Vert
 H_{\text{per}}\Vert^{2}/\sin^2(T_{c}\Delta/2)]
\label{eq:rhofactored},
\end{align}
where we have bounded the corrections due to factoring
$\exp(-i\epsilon\Omega_{N}^{\perp}) \approx
\exp(-i\epsilon\Omega_{N}^{\ket{0}})\exp(-i\epsilon\Omega_{N}^{\text{rest}})$
by $O(\epsilon^{2}\Vert\Omega_{N}^{\perp}\Vert^2)$ estimated by the
final asymptotic term. Thus, the fidelity loss is governed by
$\epsilon\Omega_{N}^{\ket{0}}$.  Let $E_{\ket{0}}=\sum_{j\ne
0,b_{1},b_{2}}(E_{0,b_{1};j,b_{2}}+\text{ h.c.})$.  Then we may bound
\[
\epsilon\Vert\Omega_{N}^{\ket{0}}\Vert\le \frac{\epsilon\Vert
E_{\ket{0}}\Vert }{ \sin(T_{c}\Delta_{\ket{0}}/2)} ,\;\;\;
\Delta_{\ket{0}}=\hspace*{-2mm}\min_{j\ne 0,b_{1},b_{2}}
\vert\omega_{j,b_{2}}-\omega_{0,b_{1}}\vert .\] 
\noindent 
Physically, the {}``relevant gap'' $\Delta_{\ket{0}}$ is the minimum
energy difference between the $\ket{0}$ sector and any other sector in
$H_D$. In principle, the latter can be refined if detailed knowledge
of the energy levels of $H_{\text{dom}}$ and $H_{B}$ is available. We
also note that $\Vert E_{\ket{0}}\Vert\le\Vert E\Vert\approx
T_{c}\Vert H_{\text{per}}\Vert$.

Using the fidelity bounds in Ref. \cite{fuchs99,Lidar-Bounds} [see
also Eq. (\ref{Bbound})], we can connect the operator norm of the
effective Hamiltonian to fidelity loss. For small $\epsilon$, we have:
\begin{align}
1-f_{N} &
 \le\epsilon\Vert\Omega_{N}^{\ket{0}}\Vert+O(\epsilon^{2}T^{2}\Vert
 H_{\text{per}}\Vert^2)+O(\epsilon^{2}\Vert
 H_{\text{per}}\Vert^{2}/\Delta^{2})\nonumber \\ &
 =\frac{\epsilon\Vert
 E^{\ket{0}}\Vert}{\sin(T_{c}\Delta_{\ket{0}}/2)}
 \label{eq:fnbound}\\
 & +O(\epsilon^{2}T^{2}\Vert
 H_{\text{per}}\Vert^2)+O(\epsilon^{2}\Vert
 H_{\text{per}}\Vert^{2}/\Delta^{2}), \nonumber
\end{align}
which we can weaken and approximate to:
\begin{equation}
f_{N}\gtrsim 1- \frac{\epsilon T_{c}\Vert
H_{\text{per}}\Vert}{\sin(T_{c}\Delta_{\ket{0}}/2)}.
\label{eq:1mfguar1}
\end{equation}
Notice that the fidelity bound in Eq. (\ref{eq:fnbound}), is similar
in structure to the semi-classical case bound from perturbation theory
in Eq. (\ref{eq:semifid}), but is a weaker bound by a power of
2. While this is due to the coarse bounds for fidelity loss in terms
of effective Hamiltonian from Ref. \cite{Lidar-Bounds}, we emphasize
that these bounds have the advantage that they apply irrespective of
the state of the environment. Also notice that the appearance of
operator bounds guarantees a {\em polynomial dependence} on the number
of subsystems $n_B$ in the environment, as all system-environment
operators can be written as sums of few-body interaction terms among
the environment subsystems.

The bound on the fidelity loss in Eq. (\ref{eq:fnbound}) is valid
provided that two conditions are met: first,
\begin{equation}
\epsilon T\Vert H_{\text{per}}\Vert\ll 1 ,
\label{eq:condB1} 
\end{equation}
which is a technical requirement for our approximations to be valid
(in particular, for neglecting $\Omega_{N}^{[2+]}$ with respect to
$\Omega_{N}^{[1]}$), and delineates the regime of applicability of MET
in our analysis; and second,
\begin{equation}
\Delta \gg \epsilon\Vert H_{\text{per}}\Vert ,
\label{eq:condB3}
\end{equation}
which is needed to ensure that no transition between the different
pointer sectors occurs due to the perturbation.  Both these
requirements can be met by reducing $\epsilon$, the perturbation
strength. Interestingly, we may interpret the gap condition in
Eq. (\ref{eq:condB3}) as a simple {\em energetic protection} of
coherence: PSs are protected indefinitely if the perturbation is too
weak to cause a major transition among them. The role of a
sufficiently gapped $H_\text{dom}$ is precisely to ensure this energy
barrier.

\section{Constructive Control Protocols }
\label{sec:Control-Sequences}

The results of the previous section provide a quantitative estimate on
the {\em maximal} fidelity loss that PSs may incur, no matter how the
latter happen to be produced.  Ideally, the system starts with a
nearly-perfect preparation (by means of a suitable quantum operation)
in one of the PSs, thus the fidelity $f_{0}=1$. As the control cycles
are repeatedly applied, the fidelity will decay quickly for an initial
transient {[}see Appendix \ref{sec:Appendix:Initial-Decay}{]}, but
will eventually enter a ``saturation'' regime with a fidelity level
lower-bounded by Eq.  (\ref{eq:1mfguar1}) once the number of applied
cycles $N$ is sufficiently large for the MET-averaging to kick in: 
\[
NT_{c}\Delta=T\Delta \gg 1. \]
\noindent 
This saturation regime will hold as long as the conditions in
Eqs. (\ref{eq:condB1})-(\ref{eq:condB3}) are satisfied.  Within this
picture, to engineer an arbitrary set of PSs, we need to:

$\bullet$ PS1: {\em Engineer the dominant cycle Hamiltonian}
$H_{\text{dom}}$ into the diagonal form of Eq. (\ref{eq:eigexp}), so
that the PS condition holds for the desired choice of
$\{\ket{j}\}_{j=0}^{p-1}$, with $1 \leq p \leq D_S -1$;

$\bullet$ PS2: {\em Minimize the perturbation} $H_{\text{per}}$, by
enforcing high modulation rates (that is, by reducing $T_c$) and/or
employing advanced control protocols similar to those in high-order DD
\cite{Khodjasteh2004,PhysRevLett.96.100405,Uhrig2007} to get a higher
fidelity bound {[}See Sec. \ref{sub:Higher-Order-Sequences}{]}.
Notice that this also implies that Eqs. (\ref{eq:condB1}) and
(\ref{eq:condB3}) can be better satisfied and thus the desired PS set
preserved to a \emph{longer }guaranteed time.

In this Section, we describe how the tasks PS1 and PS2 can be achieved
using unitary control pulses, for an arbitrary open-system setting in
which $H_{SB}$ and $H_{B}$ are quantitatively unspecified (bounded)
operators on ${\cal H}_S \otimes {\cal H}_B$.  Although in principle
continuous modulation schemes could be envisioned for this purpose,
the pulsed control setting we focus on has the advantage to be
mathematically straightforward while providing interesting connections
to both DD theory and recent experiments.  We begin by assuming that
$H_{SB}$ is {\em generic} -- that is, it has {\em no} special symmetry
hence no degeneracies -- so that the engineered PSs and only those
remain at high fidelity.  We revisit this assumption in
Sec. \ref{sub:De-Symmetrization}.

\subsection{Generation of the Effective Hamiltonian}
\label{sub:Transformation-of-Bare}

The control cycles in our construction are pulse sequences in which
free evolution intervals of duration $\tau_{i}$ are punctuated by
application of \emph{system} unitary operators $P_{i}$, so that
$T_{c}=\sum_{i=1}^n \tau_{i}$.  The operators $P_{i}$ are assumed to
be implemented as (nearly) instantaneous pulses by
$H_{\text{ctrl}}(t)$. While in this section as well as in the examples
of Sec. \ref{sec:Examples} we will also assume each pulse to be
implemented perfectly, it turns out that such ideal-pulse assumptions
are not essential as long as the unitary propagator does not change
from cycle to cycle {[}See Sec. \ref{sec:imperfect}{]}.  Similar to
DD, the control pulses are so designed that $P_{i}$ cancel each other:
$P_{n}\cdots P_{1}=I_{S}$.  The overall unitary propagator for the
cycle is an ordered product of pulse unitaries interlaced with
evolution propagators $\exp(-i\tau_{j}H_0)$ associated with the free
intervals in which $H_0=H_{SB}+H_{B}$. That is,
\begin{eqnarray} 
U_{c}(T_c) & = & P_{n}\exp(-i\tau_{n}H_0)\cdots
P_{1}\exp(-i\tau_{1}H_0)
\label{eq:ucpi}
\\ & = & Q_{n}^{\dagger}\exp(-i\tau_{n}H_0)Q_{n}\cdots
Q_{1}^{\dagger}\exp(-i\tau_{1}H_0)Q_{1}\nonumber \\ & = &
\exp(-i\tau_{n}Q_{n}^{\dagger}H_0Q_{n})
\cdots\exp(-i\tau_{1}Q_{1}^{\dagger}H_0Q_{1}) \nonumber \\ & \equiv &
\exp(-iT_{c}H_{c}),\nonumber
\end{eqnarray} 
where $Q_{j}=P_{j-1}\cdots P_{1}$ for $n\ge j>1$ and $Q_{1}=I_{S}$.
The effective cycle Hamiltonian $H_{c}$ [Eq. (\ref{eq:ucyc})] can be
approximated using a Magnus expansion:
\begin{equation}
T_{c}H_{c}=\sum_{j=1}^{n}\tau_{j} Q_{j}^\dagger H_0 Q_{j}
+O(T_{c}^{2}H_0^{2}).
\label{eq:magcyc}
\end{equation}
In the special case of a uniform sequence, for which $\tau_i\equiv
\tau= T_c/n$, we will simply denote the cycle in Eq. (\ref{eq:ucpi})
by
\[f P_1 f P_2\cdots f P_n, \]
where operations are now applied left-to-right and $f$ stands for a
free evolution interval.

The task PS1 then reduces to casting $H_{c}$ in the form of
Eqs. (\ref{eq:ucyc})--(\ref{eq:eigexp}) for a designated PS set
$\{\ket{j}\}_{j=0}^{p-1}$.  Note that in {\em non-selective}
(universal) DD schemes \cite{Viola1999Dec}, a fixed choice of $P_{i}$
(determined by the system dimensionality $D_{S}$ and the symmetry of
$H_{SB}$) is used to cancel $H_{0}$ (modulo pure-bath terms) up to
various orders in $\Vert T_c H_0\Vert$.  In this sense, PS1 is easier
than DD, as it requires cancellation of far fewer terms in $H_{c}$. In
fact, the minimum number of intervals in a cycle designed to
completely cancel a generic $H_{SB}$ is $D_{S}^{2}$
\cite{Viola1999Dec,wocjan2002}, while in contrast, as we shall prove
below, the cycle for achieving PS1 for a complete pointer basis
requires only $D_{S}$ time slots.  Achieving PS1 (and PS2) is, rather,
close in spirit to {\em selective} DD, the price to be paid, however,
being that the required control operations become necessarily {\em
state-dependent}.  A similar idea has been explored in
Refs. \cite{PhysRevLett.96.100405,Mukhtar2010,PhysRevA.82.052338},
where ``rotated'' DD pulses are used to ``lock'' states in a
particular one- or two- dimensional subspace of a two-qubit
dissipative system.  Note that selective (and encoded) DD for
subspaces \cite{VKL02,Wu-LEO} has also been specifically invoked for
dynamical quantum error suppression, in alternative to non-unitary
control strategies based on error-correcting codes \cite{Knill-QEC}.
A basic difference between the PS-engineering protocols that we will
provide and the above-mentioned methods is the \emph{dimension} of the
preserved structures. Here, the preserved states correspond to {\em
isolated points} (zero-dimensional manifolds) in the system Hilbert
space, while by decoupling subspaces (or subsystems), the
corresponding IPS corresponds to continuous manifolds of higher
dimension.

As mentioned, the desired set of PSs may comprise any number $p$ of
orthonormal states. For clarity, we start with the task of engineering
a single PS, follow with preserving a complete pointer basis, and
finally provide a general recipe for engineering an incomplete pointer
set that includes the former protocols as special cases.

\subsubsection{Single Pointer State}
\label{sub:Single-Pointer-State}

Let $\ket{0}$ be the solitary target PS ($p=1$). Define the unitary
reflection operator
\begin{equation}
Q=2\ket{0}\!\bra{0}-I_S.
\label{eq:qdef}
\end{equation} 
Notice that despite the simple representation, implementing $Q$ need
not be simple, and we further comment on that in Appendix
\ref{sec:appendix:reflect}.  Consider a uniform pulse sequence
consisting of two $Q$ pulses separated by equal intervals of duration
$\tau=T_{c}/2$. Using Eq. (\ref{eq:magcyc}), one can verify that
$H_{c}$ for this sequence obeys the PS-condition in
Eq. (\ref{eq:eigexp}),
\[
H_{c}=\ket{0}\!\bra{0}\otimes B_{0}+H_{\text{per}}+I_{S}\otimes
H_{B},\] 
\noindent 
where $B_{0}$ is a bath operator and $H_{\text{per}}=O(\tau\Vert
H_0\Vert^{2})$ \cite{PhysRevLett.96.100405,Mukhtar2010}.  Since the
latter can be made smaller by using shorter pulse intervals, the
sequence will preserve $\ket{0}$ with an arbitrarily high fidelity
that can be maintained for a long time. Thus, PS1 and PS2 are both
achieved by the control cycle $fQfQ \equiv QQ$.

If $D_S=2$, the choice of $\ket{0}$ determines a unique orthonormal
state $\ket{1}$ and Eq. (\ref{eq:qdef}) yields $Q=\ket{0}\langle 0|
-\ket{1}\langle 1| \equiv \sigma_{z}$, resulting in a sequence that we
shall simply denote $ZZ$ henceforth.  In a DD problem, this sequence
is designed to suppress terms of the form $\sigma_x\otimes B_{X}+\sigma_y\otimes
B_{Y}$ in $H_{SB}$.  In contrast, a universal DD sequence (such as
$XYXY$) is designed to cancel every possible term in $H_{SB}$,
including $\sigma_z\otimes B_{Z}$.  This generally results in an enhanced
quantum memory where \emph{all} states are preserved with a higher
fidelity \emph{initially}, but need not lead to a fidelity which is
\emph{maintained} for any particular state: even approximate
eigenstates induced by $XYXY$ are not apparent. Despite this
distinction, in Sec. \ref{sub:Qubit-System} we will analyze the
structure of $H_{SB,c}$ for both sequences in detail, and show how the
states $\ket{0}$ and $\ket{1}$ form a (accidental) pointer basis under
the $XYXY$ sequence as well.

\subsubsection{Complete Pointer Basis}
\label{sub:Complete-Pointer-Basis}

Let $\{\ket{j}\}_{j=0}^{D_{S}-1}$ denote the set of $p=D_{S}$ states
we desire to preserve. Following \cite{wocjan2002}, let us define the
unitary operator $\sigma_{D_{S}}$ as
\begin{equation}
\sigma_{D_{S}}=\sum_{j=0}^{D_{S}-1}\omega^{j+1}\vert
j\rangle\negmedspace\langle j\vert,\quad\omega=e^{2\pi i/D_{S}}.
\label{eq:tds}
\end{equation} 
Direct calculation shows that for $0\le i,j\le D_{S}-1$,
$$ \sum_{k=0}^{D_{S}-1}(\sigma_{D_{S}})^{-k}(\vert
i\rangle\negmedspace\langle j\vert\otimes
B_{ij})(\sigma_{D_{S}})^{k}=D_{S}\delta_{i,j}\vert
i\rangle\negmedspace\langle i\vert\otimes B_{ii}.$$ 
\noindent 
Now consider a control cycle consisting of equal free intervals of
length $\tau=T_c/D_S$, during which $\sigma_{D_{S}}$ is applied
$D_{S}$ times, or $f\sigma_{D_S}\cdots f\sigma_{D_S}$.  Thus,
$P_{j}=\sigma_{D_{S}}$ and $Q_{j}=(\sigma_{D_{S}})^{j}$.  Using
Eq. (\ref{eq:magcyc}), the effective cycle propagator again takes the
desired form of Eq. (\ref{eq:eigexp}):
\[ H_{c}=\sum_{j=0}^{D_{S}-1}\ket{j}\!\bra{j}\otimes
B_{j}+H_{\text{per}}+I_{S}\otimes H_{B},\] 
\noindent 
where, as before, $\Vert H_{\text{per}}\Vert=O(\tau\Vert
H_0\Vert^{2})$. 

Note that the above sequence is by no means \emph{uniquely} suited for
the tasks PS1 and PS2. For example, sequences based on Hadamard arrays
\cite{Jones1999322,Leung02} or products of $\sigma_{z}$ Pauli matrices
\cite{WangLiu2010} can also be envisioned when the system is a
collection of $n_q$ qubits ($D_S=2^{n_{q}}$). Nonetheless, the
sequence of $\sigma_{D_{S}}$ operators described here is notable since
it requires a \emph{single pulse type}. As we shall see in
Sec. \ref{sub:Imperfect-Pointer}, this implies an extra degree of
robustness against systematic control imperfections.

\subsubsection{Incomplete Pointer Basis}
\label{sub:Incomplete}

Let $\{\ket{j}\}_{j=0}^{p-1}$ denote the set of $p<D_{S}-1$
orthonormal states we wish to preserve (the case $p=D_{S}-1$ is
equivalent to $p=D_{S}$, as already noted.  The set
$\{\ket{j}\}_{j=0}^{p-1}$, possibly along with other orthonormal
states $\{\ket{j}\}_{j=p}^{D_{S}-1}$, forms an orthonormal pointer
basis for ${\cal H}_S$. The construction described in the previous
subsection can be readily modified to this scenario by introducing the
following unitary pulse operator:
\begin{equation}
\sigma_{D_{S}}^{p}=\sum_{j=0}^{p-1}\omega^{j+1}\vert
j\rangle\negmedspace\langle j\vert+\sum_{j=p}^{D_{S}-1}\vert
j\rangle\negmedspace\langle j\vert,\ \omega=e^{2\pi i/(p+1)} .
\label{eq:sigmapd}
\end{equation} 
The 1st sum in Eq. (\ref{eq:sigmapd}) corresponds to a diagonal block
of size $p$ while the 2nd sum implies that $\sigma_{D_{S}}^{p}$ acts
as identity on $\text{Span}\{\ket{j}\}_{j=p}^{D_{S}-1}$.
Consider a control cycle consisting of equal free intervals of length
$\tau=T_c/(p+1)$, during which $\sigma_{D_{S}}^{p}$ is applied $p+1$
times, or $f\sigma_{D_{S}}^{p}\cdots f\sigma_{D_{S}}^{p}$.  Using
Eq. (\ref{eq:magcyc}) with $Q_{j}=(\sigma_{D_{S}}^{p})^{j}$ for
$j=1,\dots,p+1$, we can verify that the effective cycle propagator for
this sequence obeys the PS-condition in Eq. (\ref{eq:eigexp}):
\[
U_{c}=\exp[-iT_{c}(\sum_{j=0}^{p-1}\ket{j}\!\bra{j} \otimes
B_{j}+H^{\prime}+H_{\text{per}}+I_{S}\otimes H_{B})],\]
\noindent 
where, as required, $H^{\prime}$ annihilates
$\text{Span}\{\ket{j}\}_{j=0}^{p-1}$ and again $\Vert
H_{\text{per}}\Vert=O(\tau\Vert H_0\Vert^{2})$.

\subsection{Over-Symmetric Systems}
\label{sub:De-Symmetrization}

In essence, PS engineering is about introducing and maintaining
symmetries in the open-system Hamiltonian $H_0$, and indeed the
protocols presented in Sec. \ref{sub:Transformation-of-Bare}, in
conjunction with the error bounds of
Sec. \ref{sec:Ergodic-Averaging-Induced}, succeed at achieving this
goal if $H_{SB}$ is generic, that is, not \emph{over-symmetric}. The
occurrence of {\em unwanted symmetries} can cause two distinct
problems.  First and most importantly, they can result in energy
degeneracies in $H_\text{dom}$.  This will affect the perturbation
theory requirements in our derivation of fidelity of PSs, since
vanishing energy differences [$\omega_{j}$ and $\omega_{j,b}$ in
Eqs. (\ref{eq:semifid}) and (\ref{eq:1mfguar1}), respectively] will
lead to a divergence in fidelity loss.  In such cases, obliviously
applying the above control procedures may result in PSs that decay
quickly.  Second, another problem arises when a state {\em other} than
the designated set $\{\vert j\rangle\}_{j=0}^{p-1}$ happens to satisfy
the PS condition in Eq. (\ref{eq:eigexp}) due to additional
symmetry. The intended PSs will then be preserved, but not exclusively
so.  In particular, decoherence-free subspaces or more general IPSs
\cite{Lidar-DFS,Knill-NS} could exist/emerge in the presence of the
control sequence that is used for engineering PSs.

Both the problematic scenarios of energy degeneracy and
non-exclusivity of PSs can be remedied by modifying the control
protocol so that $H_{0}$ (and consequently $H_c$) is suitably
``de-symmetrized''. In practice, a situation of unconditional failure
of the protocols described in Sec. \ref{sub:Transformation-of-Bare}
hints at hidden symmetries in the bare $H_{SB}$ that persist in the
effective cycle Hamiltonian $H_{SB,c}$, and need to be addressed by
modifying the control protocol so that the transformation from $H_{0}$
to $H_c$ not only {\em introduces} the desired symmetry but also {\em
removes} the undesired ones. A straightforward way in which such a
desymmetrization can be obtained is to ``turn on'' an constant
Hamiltonian on the system. For degenerate cases, an explicitly
diagonal and non-degenerate control Hamiltonian can be applied,
whereas for the undesired PSs an explicit coupling to yet another
state (if available) will guarantee the desired desymmetrization.
Another (pulsed) solution is to adjust the previous control protocols
to remove the over-symmetry, as we describe next.

For simplicity, let us focus on a simple case where a single state
$\vert s\rangle$ is the source of the problem: it either shares energy
eigenvalues with another fellow orthogonal PS, or it is an undesired
PS.  We will assume that another state $\vert t\rangle$, orthogonal to
$\vert s\rangle$ and the rest of the PSs, exists and that the
following unitary (self-inverse) operator,
\[  R=\vert s\rangle\!\langle t\vert+\vert t\rangle\!\langle
s\vert+\sum_{i\ne s,i\ne t}\vert i\rangle\!\langle i\vert,
\]
is available for control. Notice that $R$ permutes the states $\vert
s\rangle$ and $\vert t\rangle$. Let the control cycle for generating
the desired PSs [Sec. \ref{sec:Control-Sequences}] be given by
$fP\cdots fP$, where each pulse is originally applied $n$ times. We
will ammend this sequence by concatenating \cite{Khodjasteh2004} it with $fRfR$, that
is,
\[
[fRfR]\ P[fRfR]\ P\cdots [fRfR]\ P.
\]
As a result, the number of pulses is now $2n$ and they alternate
between $RP$ and $R$. Using Eq. (\ref{eq:magcyc}), the cycle
Hamiltonian becomes:
\begin{eqnarray*}
H_{SB,c} = \sum_{k=0}^{p-1}P^{-k}(RH_{SB}R+H_{SB})P^{k},
\end{eqnarray*}
in which the inner expression $(RH_{SB}R+H_{SB})$ effectively removes
the degeneracy by ``mixing'' the states $\vert s\rangle$ and $\vert
t\rangle$ as long as there is no further symmetry among them. Notice
that, if the symmetry of $H_{SB}$ is manifest, such as in a system
exposed to collective noise, the operator $R$ can be chosen from the
outset to be a product of \emph{unequal} unitary operations on each
subsystem, which inevitably results in desymmetrization.


\section{Illustrative Applications}
\label{sec:Examples}

\subsection{Single-Qubit Sequences}
\label{sub:Qubit-System}

We begin by considering the task of engineering two arbitrary orthogonal
states $\ket{0}$ and $\ket{1}$ as a pointer basis in a qubit.  This
basis defines  the relevant Pauli operators, and the bare
system-bath interaction $H_{SB}$ can be correspondingly expanded as
\[ H_{SB}=\sigma_{x}\otimes B_{x}+\sigma_{y}\otimes
B_{y}+\sigma_{z}\otimes B_{z} , \]
\noindent 
for generic (traceless) operators $\{B_{\alpha}\}$.  As before, let
$H_{B}$ be the bath internal Hamiltonian.  Our goal is to thus
preserve the eigenstates $\ket{0}$ and $\ket{1}$ of $\sigma_{z}$ and,
as mentioned, we shall both analyze the $ZZ$ PS-protocol and the
$XYXY$ universal DD sequence.  We can approximate the effective cycle
Hamiltonian $H_c$ using the Magnus expansion.  For the $ZZ$ sequence
we have $T_{c}=2\tau$ and
\begin{eqnarray} 
H_{\text{dom}}^{[ZZ]} & = & \sigma_{z}\otimes B_{z},
\label{eq:hzzdom}
\\ \epsilon H_{\text{per}}^{[ZZ]} & = &
\frac{\tau}{2}[\sigma_{x}\otimes B_{x}+\sigma_{y}\otimes
B_{y},\sigma_{z}\otimes B_{z}+H_{B}]\nonumber \\ & + & O(\tau^{2}).
\label{eq:hzzper}
\end{eqnarray} 
The perturbative parameter in this setting is clearly proportional to
$\tau\Vert B_{\alpha}\Vert$, where $\alpha=x,y$.  For the $XYXY$
sequence we have $T_{c}=4\tau$ and
\begin{eqnarray} 
H_{\text{dom}}^{[XYXY]} & = &
4\tau \Big( 2i\sigma_{x}\otimes[B_{x},H_{B}]\nonumber \\ & + &
\sigma_{z}\otimes(i[B_{z},H_{B}]+\{B_{y},B_{x}\}) \Big),
\label{eq:xyxyeq}
\end{eqnarray}
whilst $\Vert H_{\text{per}}^{[XYXY]}\Vert=O(\tau^{2})$.  For both
sequences, higher-order corrections in $\epsilon H_{\text{per}}$ will
generically contain pure-bath terms, causing $H_{B,c}$ to differ from
the bare $H_B$.

Clearly, $H_{\text{dom}}^{[ZZ]}$ has separable eigenstates of the form
$\ket{i}\otimes\ket{b_{i}}$ for $i=0,1$. For
$H_{\text{dom}}^{[XYXY]}$, these eigenstates appear independently of
$B_x$ only if $H_B=0$, that is, the bath is non-dynamical.
Physically, the non-dynamical bath is a special but important case
which provides, in particular, a prevalent approximation for
decoherence of localized electronic spins in semi-conductors
\cite{Wen01,Wen08}.  Interestingly however, even for a generic {\em
dynamical} bath, $\ket{0}$ and $\ket{1}$ are still preserved under the
$XYXY$ sequence.  To see this we have to redefine
$H_\text{dom}$. Recall that the fidelity preservation of PSs relies on
the averaging caused by $H_D$, which since $H_\text{dom}$ is made
arbitrary small by shrinking $\tau$, is dominated by $H_B$. Consider
the bath-operators appearing in $H_{\text{dom}}^{[XYXY]}$ in
Eq. (\ref{eq:xyxyeq}).  The key observation is to realize that for any
operator of the form $[B_\alpha, H_B]$, the component along the
commutant of $H_B$ vanishes (as long as $B_\alpha$ is traceless), and
hence these terms will act solely as perturbations according to
Eq. (\ref{residual}). In contrast, the anti-commutator
$\{B_{y},B_{x}\}$ will typically have a significant component along
the commutant of $H_B$, and will break the symmetry in favor of the
$\sigma_z$ eigenstates. Thus, we may redefine the dominant Hamiltonian
responsible for PSs as
\begin{equation}
H_{\text{dom}}^{[XYXY],\Vert}=4\tau\sigma_{z}
\otimes\{B_{y},B_{x}\}^{\Vert},
\label{eq:xyxymet}
\end{equation}
which clearly still preserves $\ket{0}$ and $\ket{1}$.  A quantitative
analysis of the resulting long-time fidelity follows.

\subsubsection{Semi-Classical Environment: Analytical Results}
\label{sub:Classical-Environment}

The operators $H_{\text{dom}}^{[XYXY]}$ and $H_{\text{dom}}^{[ZZ]}$
can be calculated analytically within the semi-classical approximation
of a static random-field environment described in
Sec. \ref{sub:Semi-Classical-Bath}.  Phenomenologically, such a random
magnetic field $\mathbf{b}=(b_{x},b_{y},b_{z})$ can approximate the
effect of a nuclear spin environment interacting with a central spin
qubit system under ensemble measurements \cite{merkulov2002}.  In
addition, as we shall describe in Sec. \ref{sec:imperfect}, systematic
pulse imperfections can also formally mimick a classical always-on
magnetic field.  By invoking Eq. (\ref{eq:fNhxhyhz}) with
$h_z(\mathbf{b})=b_z$, we obtain the ensemble-averaged fidelity loss:
\begin{equation}
\langle 1-f_N\rangle_{\mathbf{b}}=\epsilon^{2}\hspace*{-1mm}\int
\hspace*{-0.5mm}\frac{\left[h_{x}(\mathbf{b})^{2}+h_{y}(\mathbf{b})^{2}\right]
\sin^{2}\left(NT_{c}b_{z}\right)}{4
b_{z}^{2}}P(\mathbf{b})d\mathbf{b},
\label{eq:qubitcl}
\end{equation}
where $P(\mathbf{b})d\mathbf{b}$ is the probability density associated
with the distribution of $\mathbf{b}$. In the limit of
$N\rightarrow\infty$, the rapidly oscillating term
$\sin^{2}\left(NT_{c}b_{z}\right)$ is smoothed out into a saturating
(plateau) behavior. In contrast, for states other than the
$\sigma_{z}$-eigenstates, the error grows to a {\em maximal} (unit)
fidelity loss value.

It is illustrative to compute the average fidelity loss for the $ZZ$
sequence with a specific distribution for $\mathbf{b}$. The
calculation carries over to other sequences and can easily incorporate
models of pulse imperfection reflected in different distributions for
$\mathbf{b}$ {[}see Sec. \ref{sec:imperfect}{]}.  Let us assume,
specifically, that the distribution is isotropic and the magnitude $b$
is distributed according to a normal distribution $P(b)$ with zero
mean and standard deviation $B$.  The expressions for
$h_{x}(\mathbf{b})$ and $h_{y}(\mathbf{b})$ can be read off
Eq. (\ref{eq:hzzper}).  The average fidelity loss can thus be
calculated in a straightforward manner.  For small $N$, we have:
\begin{equation}
\langle1-f_N\rangle_{\mathbf{b}}=\frac{2}{5}B^{4}N^{2}\tau^{4}
+O\left(N^{3}\right)\approx\frac{1}{10} B^{4}T^{2}\tau^{2}.
\label{eq:semicshort}
\end{equation}
In this regime, the fidelity decays quadratically with both the
elapsed time $T$ and the control time scale $\tau$, also in line with
general error bounds for cyclic DD \cite{Viola2005Random}.  In
contrast, for large $N$, the average fidelity loss saturates to a
limiting (time-independent) value:
\begin{eqnarray} 
\langle1-f_\infty\rangle_{\mathbf{b}} =
\frac{B^{2}\tau^{2}}{12}+O(e^{-8 B^{2}N^{2}\tau^{2}}
\hspace*{-0.5mm}N^{-2}) \approx\frac{B^{2}\tau^{2}}{12},
\label{eq:semiclong}
\end{eqnarray}
which is controlled by the product $B\tau$.  The pattern of initial
fidelity decay, followed by saturation, is the same pattern observed
in exact numerical simulations of DD in quantum dots
\cite{Wen01,Wen08}.  In such a setting, the bath consists of a large
number $n_{B}$ of nuclear spins, each coupled to the central spin with
a strength $j_m$ and with
$$A \equiv \Big( \sum_{m=1}^{n_B} \frac{j_{m}^{2}}{n_B} \Big)^{1/2}$$
\noindent 
being a measure of the coupling strength {[}see also the upcoming
Eq. (\ref{eq:jmsi}){]}. In the semi-classical limit, the environmental
spins produce a classical Overhauser field.  Using this analogy and
the exact result for the asymptotic fidelity derived in
Ref. \cite{Wen01,Wen08} for a maximally mixed bath initial state, that
is, $1-f_\infty = \tau^2 A^2 n_B/16$, we can interpret the variance of
the $B$-field as follows:
\begin{equation} 
B^{2}=\frac{3}{4} A^{2}n_{B}.
\label{eq:b2}
\end{equation} 
Thus, we recover the polynomial dependence of the long-time fidelity
of the engineered qubit PS's on $n_{B}$, as argued in
Sec. \ref{sub:Toggling-Frame-of} on general grounds. This will be also
verified in the upcoming numerical results
{[}Figs. \ref{fig:errorsattau}-\ref{fig:nbdep}{]}.

The initial decay of fidelity followed by a saturation (or freeze)
regime is consistent with the more general semi-classical analysis
carried out in \cite{prosen2003,prosen2005,Weinstein-Eigen}.  The
semi-classical approach also highlights an interesting connection to
the Krylov-Bogoliubov method of averaging in classical dynamics
\cite{krylov1949introduction}, where an oscillatory motion is replaced
by an average that yields approximate integrals of motion.  In the
quantum treatment, a similar behavior emerges from the theoretical
model developed in Sec. \ref{sub:Toggling-Frame-of}, whereby the
initial decay is followed by a bounded fidelity loss in the MET
regime. Note that unlike the classical limit, after many oscillations
(possibly when $T$ no longer satisfies Eqs. (\ref{eq:condB1}) or
(\ref{eq:condB3})), fidelity may eventually drop, as the linear
approximation implicit in the MET theorem (ignoring $\Omega^{[2+]}$
terms) need no longer hold.

\subsubsection{Quantum Environment: Exact Numerical Results}

\begin{figure*}
\begin{centering}
\includegraphics[width=0.85\textwidth]{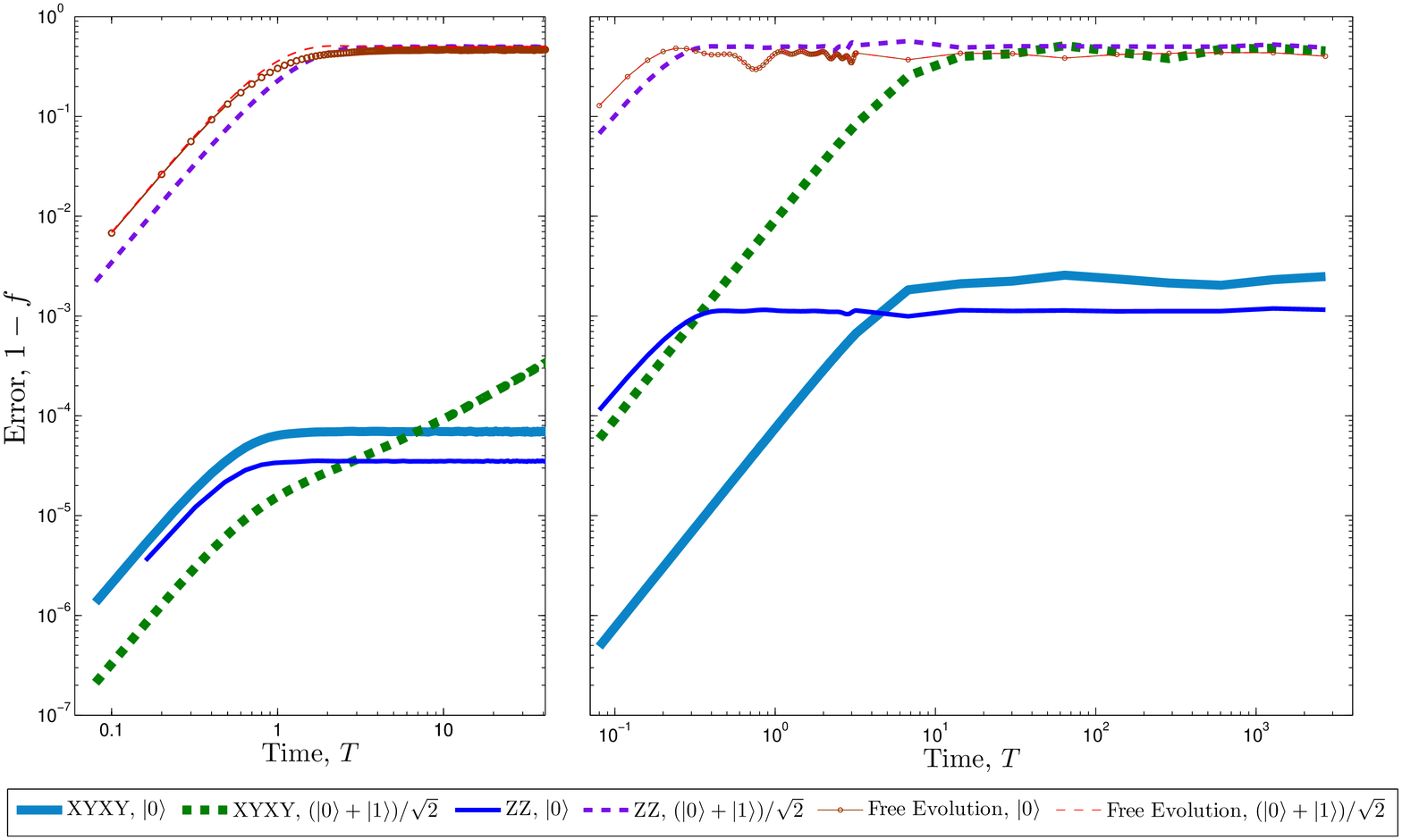} \par\end{centering}
\centering{}\caption{(Color online) Fidelity error ($1-f$) as a
function of time for $XYXY$ and $ZZ$ sequences as well as the free
evolution for $\ket{+X}=(\ket{0}+\ket{1})/\sqrt{2}$ and
$\ket{+Z}=\ket{0}$ initial states. Left: $n_{B}=15$ bath spins. Right:
$n_{B}=8$ bath spins.  In both cases the bath is non-dynamical
($\beta=0$), and the pulse interval is set at $\tau=0.01$ in arbitrary
units.  In order to allow for a comparable effective $B$-field
[Eq. (\ref{eq:b2})] and error range, the maximum coupling $\vert
J_L\vert = 1$ ($\vert J_R\vert = 4$) in the left (right) panel,
respectively, with time kept in units of $1/J_L$ in both cases.  The
difference in the time spans is due to practical limitations.  Note
that here and in the figures that will follow, the time axis is
sampled non-uniformly to emphasize the interesting features of each
time series without using too many data points.  The sample points are
connected by lines as a visual aid. }
\label{fig:Fidelity-loss-error}
\end{figure*}

We illustrate our findings by means of exact numerical simulations
where a spin qubit {[}$\mathbf{S}=(S_{x},S_{y},S_{z})$ spin
operators{]} couples through Heisenberg interaction terms,
\begin{equation}
H_{SB}=\sum_{m=1}^{n_{B}}j_{m}\mathbf{S}\cdot\mathbf{I}^{(m)},
\label{eq:jmsi}
\end{equation}
to a quantum environment consisting of spin-1/2 particles
[$\mathbf{I}^{(m)}$ spin operators].  The (bare) internal Hamiltonian
of the spin bath is taken to be of dipolar form:
\begin{equation}
H_{B}= \sum_{m=1}^{n_{B}}\sum_{k<m} \beta_{mk}
(I_{x}^{(m)}I_{x}^{(k)}+I_{y}^{(m)}I_{y}^{(k)}-2I_{z}^{(m)}I_{z}^{(k)}).
\label{eq:dipolar}
\end{equation}
The coupling constants $j_{m}$ between the system spin and each bath
spin are arbitrarily generated by (uniformly) randomly sampling
between $-J$ and $J$. Similarly, the coupling constants $\beta_{mk}$
between each two bath spins are randomly chosen between $-\beta$ and
$\beta$.  We also assume that initially the bath is in a fully mixed
state.

Fig. \ref{fig:Fidelity-loss-error} compares the fidelity loss of
different initial preparations under free evolution and under the $ZZ$
and $XYXY$ sequences designated to preserve $\ket{0}\equiv \ket{+Z}$
and $\ket{1}\equiv \ket{-Z}$. Specifically, the system is prepared
either as $\ket{+Z}$ (intended as a PS) or
$\ket{+X}=(\ket{0}+\ket{1})/\sqrt{2}$ (\emph{not} intended as a PS).
We observe that for the free evolution or for state $\ket{+X}$, the
fidelity loss approaches a maximal value corresponding to a completely
mixed state. Degradation is significantly faster for the free
evolution and $ZZ$ with $\ket{+X}$) and considerably slower for the
$XYXY$ (with $\ket{+X}$), as DD will nonetheless universally extend
coherence times. In fact, the state $\ket{+X}$ in the model with the
larger environment (left panel) did not reach the maximal loss state
within the simulation time under the $XYXY$ sequence.  On the other
hand, with either the $ZZ$ or $XYXY$ sequence, the desired PS
$\ket{+Z}$ is preserved with high fidelity after the initial transient
is over.  This initial decay is expected to occur before the MET
regime kicks in and is approximately described by
Eq. (\ref{eq:semicshort}).  Notice that the value of fidelity is
non-trivially bounded after the initial decay, and furthermore it
converges smoothly to a saturation value for all engineered PSs in
Fig. \ref{fig:Fidelity-loss-error}.  The saturation behavior is in
agreement with the semi-classical result in
Eq. (\ref{eq:semiclong}). We also point out that the initial fidelity
loss values depicted in Fig. \ref{fig:Fidelity-loss-error} correspond
to a the first period of control, at $T=T_c$, which is expected to
scale with $\Vert T_c H_\text{per}\Vert^2$.

\begin{figure}
\begin{centering}
\includegraphics[width=\columnwidth]{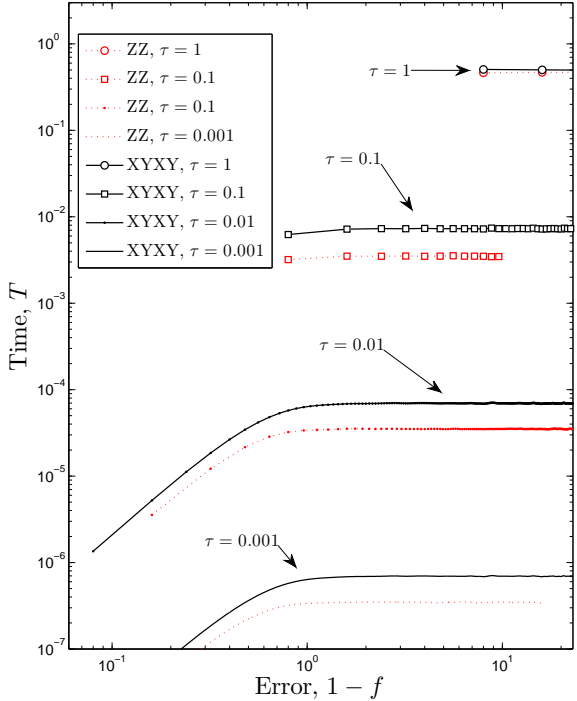}
\par\end{centering} \centering{}\caption{(Color online) Fidelity error
($1-f$) of $\ket{0}$ as a function of time for $XYXY$ and $ZZ$
sequences with different pulse intervals $\tau$.  The number of bath
spins $n_{B}=15$ and the coupling is capped at $\vert J\vert=4$.}
\label{fig:errorwithtau}
\end{figure}

Fig. \ref{fig:errorwithtau} depicts the change in fidelity of the
preserved state $\ket{0}$ by applying the sequences at different pulse
intervals $\tau$. As expected, faster modulations (smaller $\tau$)
results in a smaller perturbation strength and thus higher
fidelities. For both sequences, the ratio $\Vert \epsilon
H_{\text{per}}/H_{\text{dom}}\Vert =O(\tau)$ determines the long-time
fidelity loss {[}Eq. (\ref{eq:1mfguar1}){]}, but the difference in the
structure of $H_{\text{per}}$ vs. $H_{\text{dom}}$ in the two cases
{[}Eqs. (\ref{eq:hzzdom})--(\ref{eq:xyxyeq}){]} is responsible for the
noticeable difference between the final fidelity saturation value, the
ZZ sequence outperforming the XYXY DD protocol.

\begin{figure}[t]
\begin{centering}
\includegraphics[width=0.98\columnwidth]{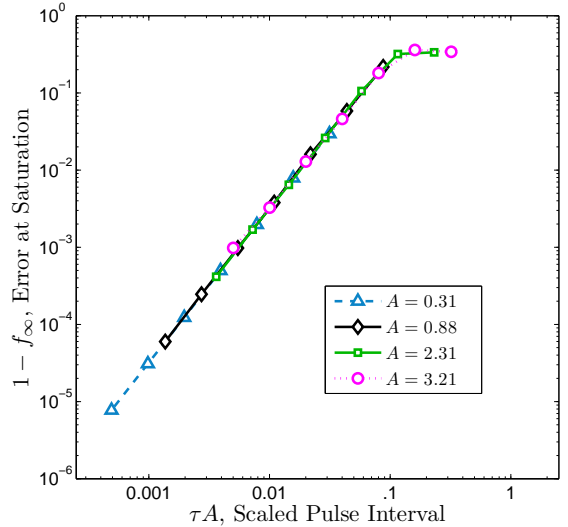}
\par\end{centering} \centering{}\vspace*{-6mm}\caption{(Color online)
Long time (saturated) fidelity error ($1-f_{\infty}$) as a function of
rescaled interval $\tau A$ for preserving $\ket{0}$ with the $ZZ$
sequence, and $A=(\sum_{m=1}^{n_{B}}j_{m}^{2}/n_{B})^{1/2}$.  The
expression $1-f_{\infty}=0.95(\tau A)^{1.99}$ {[}not shown in the
figure{]} is a power-law fit to the series with the smallest $A$.  The
number of bath spins $n_{B}=8$ and $H_{B}=0$. The overlapping curves
correspond to randomly generated coupling patterns between the system
qubit and the bath spins.}
\label{fig:errorsattau}
\end{figure}

\begin{figure}[h]
\begin{centering}
\includegraphics[width=0.98\columnwidth]{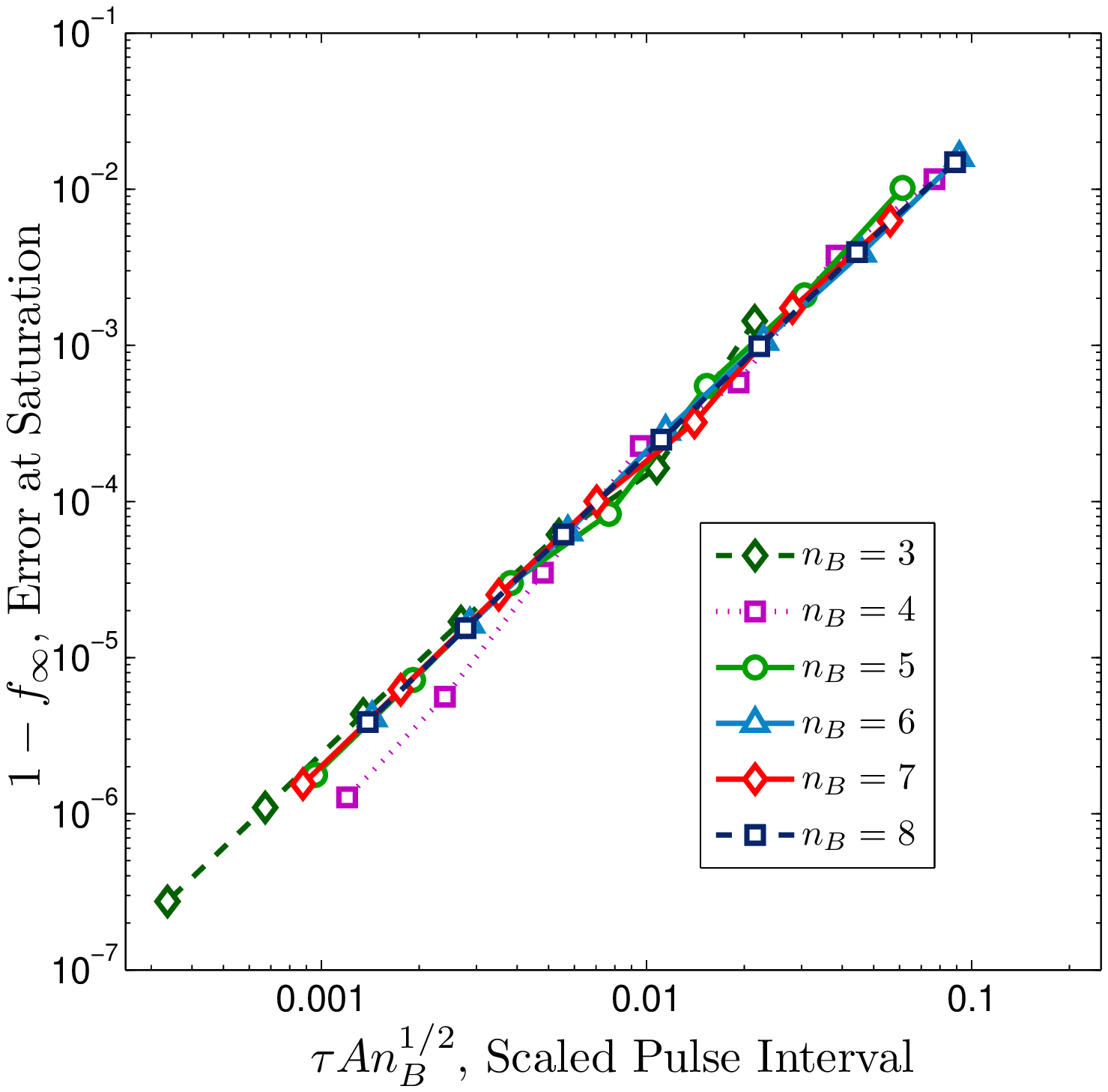} \par\end{centering}
\centering{}\vspace*{-3mm}\caption{(Color online) Long time
(saturated) fidelity error ($1-f_{\infty}$) as a function of
bath-size-rescaled interval $\tau An_{B}^{1/2}$ for preserving
$\ket{0}$ with the $ZZ$ sequence.  The simulation used
$n_{B}=3,\cdots,8$, with randomly generated couplings. The remaining
parameters are the same as in Fig. \ref{fig:errorsattau}.}
\label{fig:nbdep}
\end{figure}
 
In Fig. \ref{fig:errorsattau}, we directly probe the dependence of the
limiting fidelity ($1-f_{\infty}$) on the pulse interval $\tau$ of the
applied sequences, for various randomly generated coupling patterns
$\{j_{m}\}$ {[}Eq. (\ref{eq:jmsi}){]}. We can readily verify that the
observed fidelity loss scales with $(A\tau)^{2}$, as expected.  In
addition, Fig. \ref{fig:nbdep} superposes the fidelity saturation
values corresponding to environments of different size
($n_{B}=3,\cdots,8$), with the horizontal axis set to the
bath-size-rescaled pulse interval $A\tau
n_{B}^{1/2}$ \cite{Wen01}.  The fact that there is little variation in
the curves confirms the validity of the semi-classical approximations
leading to in Eqs. (\ref{eq:semiclong})-(\ref{eq:b2}), and provides
evidence that our simulation results may be reliably extrapolated to
realistic environments with a far larger number of spins.

\begin{figure}[h]
\begin{centering}
\includegraphics[width=0.8\columnwidth]{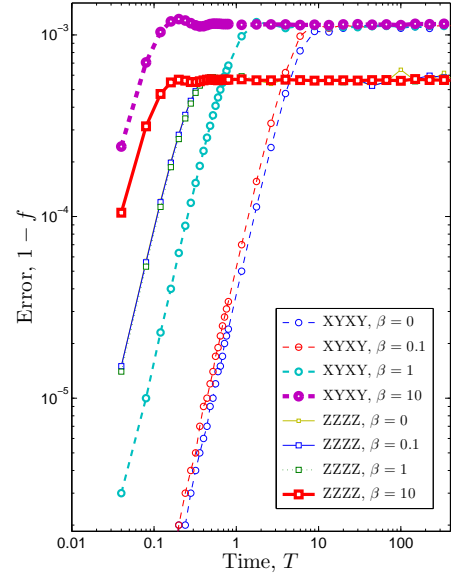}
\par\end{centering} \centering{}\vspace*{-4mm}
\caption{(Color online) Fidelity error ($1-f$) of $\ket{0}$ as a
function of time for $XYXY$ and $ZZ$ sequences with different scales
of environmental couplings $\beta$ in units of $J$. The interval is
fixed at $\tau=0.01/J$. Notice that for larger values of $\beta$ the
data series become almost indistinguishable.  The number of spin baths
$n_{B}=8$.}
\label{fig:errorwithbath}
\end{figure}

\begin{figure}[t]
\begin{centering}
\includegraphics[width=0.9\columnwidth]{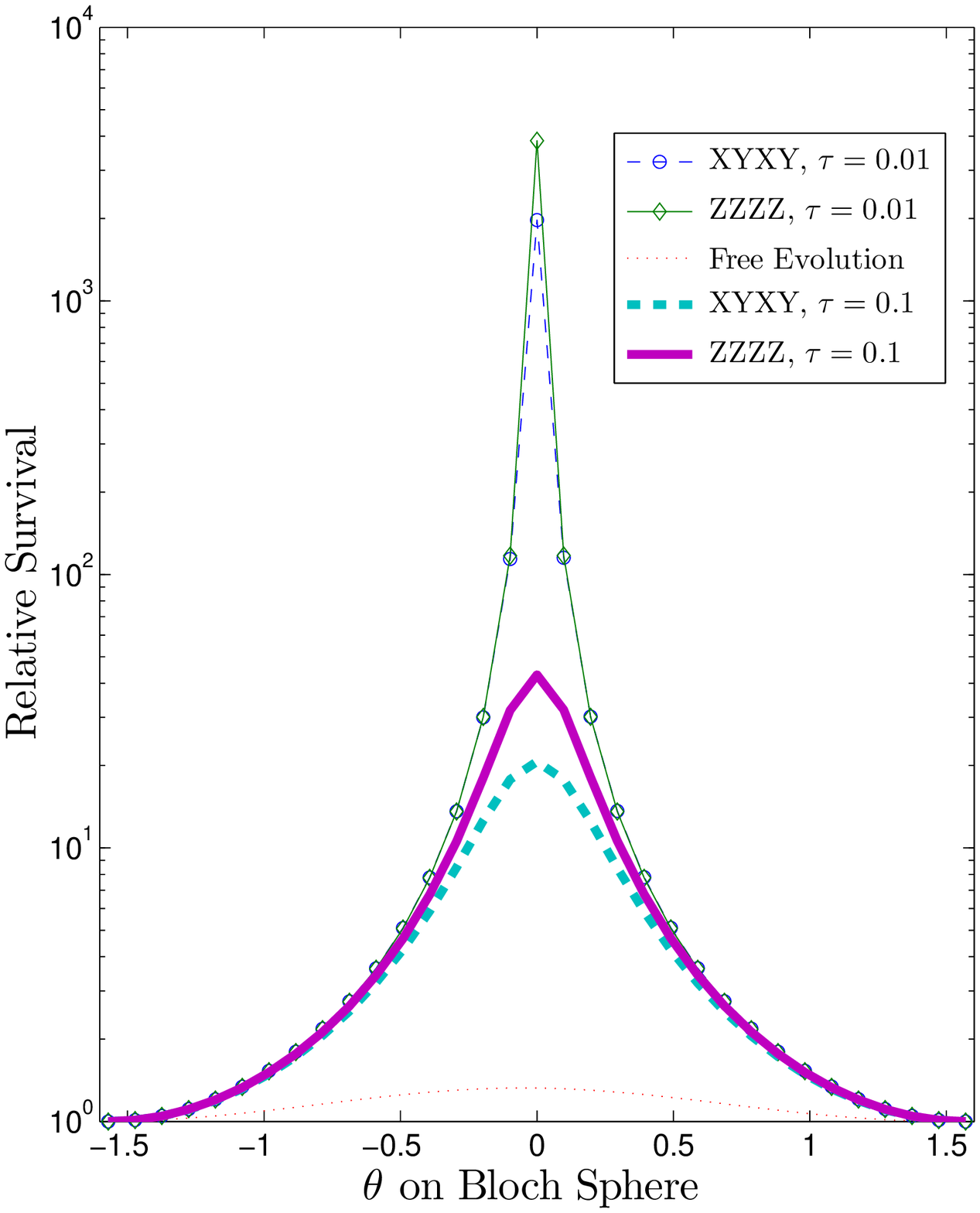}
\par\end{centering} \centering{}\vspace*{-4mm}\caption{(Color online)
Relative survival $\xi (\ket{\psi})$ (see text) as a function of time
for various sequences across the Bloch sphere equator, with
$\theta=\pm \pi/2$ corresponding to $\ket{\pm X}$.  Notice the high
survival fidelity near the pointer basis element $\ket{0}$ ($\theta
=0)$. The simulations used $n_{B}=8$ bath spins with $H_{B}=0.$
\label{fig:relsurv}}
\end{figure}

The numerical data presented thus far have addressed only
{}``non-dynamical baths'', in which the internal coupling strength
parameter $\beta=0$.  In the more general case where $\beta \ne 0$,
the fidelity saturation value shows no significant dependence on
$\beta$ over the parameter range we explored.  Results are summarized
in Fig. \ref{fig:errorwithbath}.  The independence of saturation
fidelity from $\beta$ is expected within the validity of the MET
regime as well as in the semi-classical approximation.  On the other
hand, the value of fidelity loss \emph{right after the initial cycle}
does depend on $\beta$, and more strongly so for $XYXY$ than $ZZ$.

Finally, note that for initial states other than $\ket{X+}$ that are
closer to $\ket{Z\pm}$ PSs, the fidelity loss is smaller than the
fidelity loss for $\ket{X+}$.  In Fig. \ref{fig:relsurv}, we compare
the relative survival of initial states in the $xy$-plane of the Bloch
sphere, $\ket{\psi}=\cos(\theta/2)\ket{0}+\sin(\theta/2)\ket{1}$, for
various values of $\theta$ for free evolution and the sequences $XYXY$
and $ZZ$, respectively. We have defined the {\em relative survival} of
a state as the inverse ratio of its long time fidelity loss and the
maximal fidelity loss, that is, $\xi (\ket{\psi})
\equiv (1-f_\text{min})/(1-f_\infty (\ket{\psi}))$, where
$f_{\text{min}}=1/2$ for a qubit (corresponding to the fully mixed
state).  Clearly, the engineered PSs ($\theta=0$) enjoy the highest
survival ratios for sufficiently small $\tau$'s
{[}Eq. (\ref{eq:semiclong}){]}, as intended. Also notice that
significant difference between the engineered PSs and the other states
is virtually non-existent in the free evolution.

\subsection{Bell-State Pointer Engineering}
\label{sub:EPR-Basis-Preservation}

We next consider a two-qubit spin system, coupled to a spin-1/2
environment similar to the one described in the single-qubit case.
The spin operators for the two qubits are now denoted by $S^{(1)}$ and
$S^{(2)}$, respectively.  Each qubit interacts individually with the
bath spin particles via a Heisenberg Hamiltonian, thus
\[ H_{SB}=\sum_{k=1}^{2}\sum_{m=1}^{n_{B}}j_{m,k}
\mathbf{S}^{(k)}\cdot\mathbf{I}^{(m)},\] 
\noindent 
where as before we take the $j_{m,k}$ coupling constants to be sampled
from a random distribution with $\max_{m,k}\vert j_{m,k}\vert\le J$,
and no additional symmetry is present (in particular, $j_{m,1} \ne
j_{m,2}$ for at least one $m$).  The bath internal dynamics is
governed by the dipolar interaction described by
Eq. (\ref{eq:dipolar}), where we choose $\beta=J$.  In addition, we
also allow for an always-on Heisenberg interaction to be present among
the qubits, that is,
\[ H_{S}=K\mathbf{S}^{(1)}\cdot\mathbf{S}^{(2)}.\] 
\noindent 
For concreteness, unless otherwise specified we shall match the
intra-qubit interaction strength to that between the qubits and the
environment spins, $K=J$. This corresponds to a coherent evolution
between the qubits that is comparable to (and competing with) the
interaction with the nuclear spin environment.  We will use pulse
sequences described in Sec. \ref{sec:Control-Sequences} for preserving
various sets of the Bell basis states:
\begin{eqnarray*}
\left\{ \begin{array}{l}
\ket{\text{EPR}_{0}} =  (\ket{01}+\ket{10})/\sqrt{2},\\
\ket{\text{EPR}_{1}} =  (\ket{01}-\ket{10})/\sqrt{2},\\
\ket{\text{EPR}_{2}} =  (\ket{00}+\ket{11})/\sqrt{2},\\
\ket{\text{EPR}_{3}} =  (\ket{00}-\ket{11})/\sqrt{2}.
\end{array}\right. 
\end{eqnarray*}
In particular, we engineer three sets of PSs with $p=1, 2, 4$ states
chosen from the Bell basis. In what follows, all operators are given
in the computational basis.

The sequence E1 is designed to preserve $\ket{\text{EPR}_{1}}$
(singlet state) only.  The cycle for E1 is based on the prescription
given in Eq. (\ref{eq:qdef}), with $\ket{0}$ being chosen as the EPR
basis elements. With respect to the computational basis, the required
control cycle consists then of two applications of the \textsc{swap}
gate (see also \cite{Mukhtar2010}):
\begin{eqnarray*}
U_{\text{E1}}  =  \left(\begin{array}{cccc} 1 & 0 & 0 & 0\\ 0 & 0 &
1 & 0\\ 0 & 1 & 0 & 0\\ 0 & 0 & 0 & 1\end{array}\right)
=\exp[-i\pi \mathbf{S}^{(1)}\cdot\mathbf{S}^{(2)}] e^{i\pi/4},
\end{eqnarray*}
Clearly, this operation can be implemented by the Heisenberg exchange
interaction.

\begin{figure*}[t]
\begin{centering}
\includegraphics[width=0.85\textwidth]{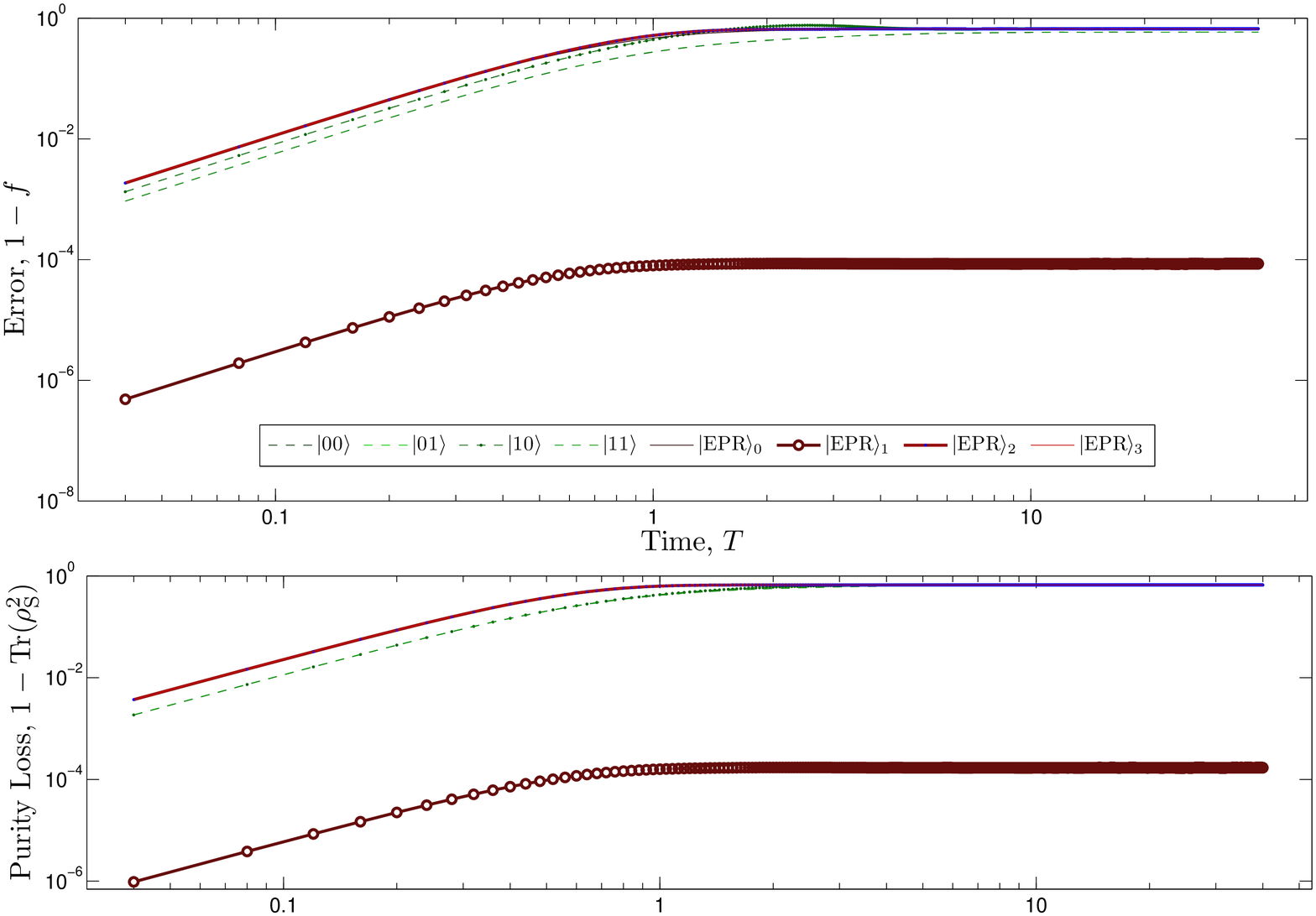} \par\end{centering}
\centering{}\caption{(Color online) Fidelity error ($1-f$) and purity
loss ($1-\text{Tr}(\rho_{S}^{2})$) for the $E1$ sequence designated to
preserve $\ket{\text{EPR}_{1}}$ (see text). Different curves (some
overlapping) correspond to initialization in the computational basis
states and the Bell basis states, respectively. The number of bath
spins $n_{B}=15$, with $\beta =J=K$, and the pulse interval
$\tau=0.01/J$.}
\label{fig:epr-1}
\end{figure*}

Similarly, the sequence E2 is designed for preserving
$\{\ket{\text{EPR}_{1}},\ket{\text{EPR}_{2}}\}$ only, and is
constructed using Eq. (\ref{eq:sigmapd}). The cycle for E2 consists of
3 applications of the unitary operation $\sigma_{p=2,D_S=4}$ which,
upon changing from the Bell to to computational basis, takes the form
\begin{eqnarray*} 
U_{\text{E2}} & = & \left(\begin{array}{cccc} \frac{1+i\sqrt{3}}{4} &
0 & 0 & \frac{-3+i\sqrt{3}}{4}\\ 0 & \frac{1-i\sqrt{3}}{4} &
\frac{3+i\sqrt{3}}{4} & 0\\ 0 & \frac{3+i\sqrt{3}}{4} &
\frac{1-i\sqrt{3}}{4} & 0\\ \frac{-3+i\sqrt{3}}{4} & 0 & 0 &
\frac{1+i\sqrt{3}}{4}\end{array}\right)\\ & = & \exp[\frac{i 4\pi}{3}
( S_{x}^{(1)}S_{x}^{(2)} + S_{z}^{(1)}S_{z}^{(2)} ) ],
\end{eqnarray*}
that is, in terms of an isotropic XZ Hamiltonian. 

Finally, the sequence E3 is designed to preserve all four EPR states
$\{\ket{\text{EPR}_{i}}\}_{i=0}^{3}$ (recall that it would be
impossible to preserve exactly 3 orthogonal states in a 4-dimensional
Hilbert space, as the remaining basis element would also be inevitably
preserved) and is constructed using Eq. (\ref{eq:tds}).  Upon
transforming, again, to the computational basis, the cycle for E3
consists of 4 applications of
\begin{eqnarray*}
U_{\text{E3}} & = & \left(\begin{array}{cccc} \frac{i-1}{2} & 0 & 0 &
\frac{i+1}{2}\\ 0 & \frac{1-i}{2} & \frac{i+1}{2} & 0\\ 0 &
\frac{i+1}{2} & \frac{1-i}{2} & 0\\ \frac{-i-1}{2} & 0 & 0 &
\frac{i-1}{2}\end{array}\right)\\ & = & \exp[{i\pi}
(S_{x}^{(1)}S_{x}^{(2)}+2S_{z}^{(1)}S_{z}^{(2)})] e^{i\pi/4},
\end{eqnarray*} 
Notice that all the sequences E1, E2, and E3 require two-body
interactions. This is unsurprising since the Bell basis is entangled.

\begin{figure*}[t]
\begin{centering}
\includegraphics[width=0.85\textwidth]{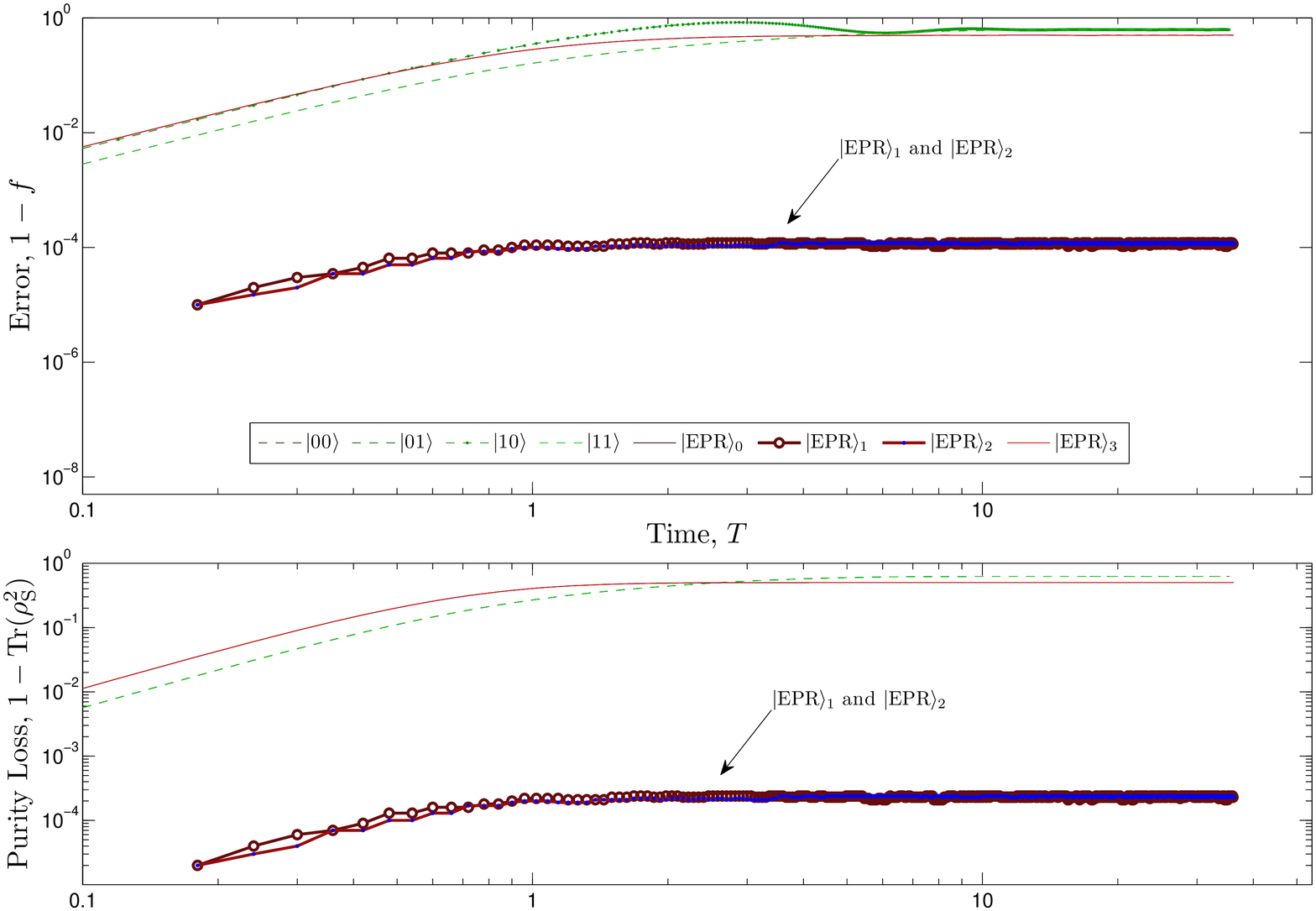} \par\end{centering}
\centering{}\caption{(Color online) Fidelity error ($1-f$) and purity
loss ($1-\text{Tr}(\rho_{S}^{2})$) for the $E2$ sequence designated to
preserve $\{\ket{\text{EPR}_{1}},\ket{\text{EPR}_{2}}\}$ (see
text). Different curves (some overlapping) correspond to
initialization in the computational ba-sis states and the Bell basis
states, respectively.  The number of bath spins $n_{B}=15$, with
$\beta =J=K$, and the pulse interval $\tau=0.01/J$.  The apparently
missing short-time fidelity/purity values for
$\{\ket{\text{EPR}_{1}},\ket{\text{EPR}_{2}}\}$ are too close to $0.0$
to be shown on the log scale.}
\label{fig:epr-12}
\end{figure*}

\begin{figure*}[t]
\begin{centering}
\includegraphics[width=0.85\textwidth]{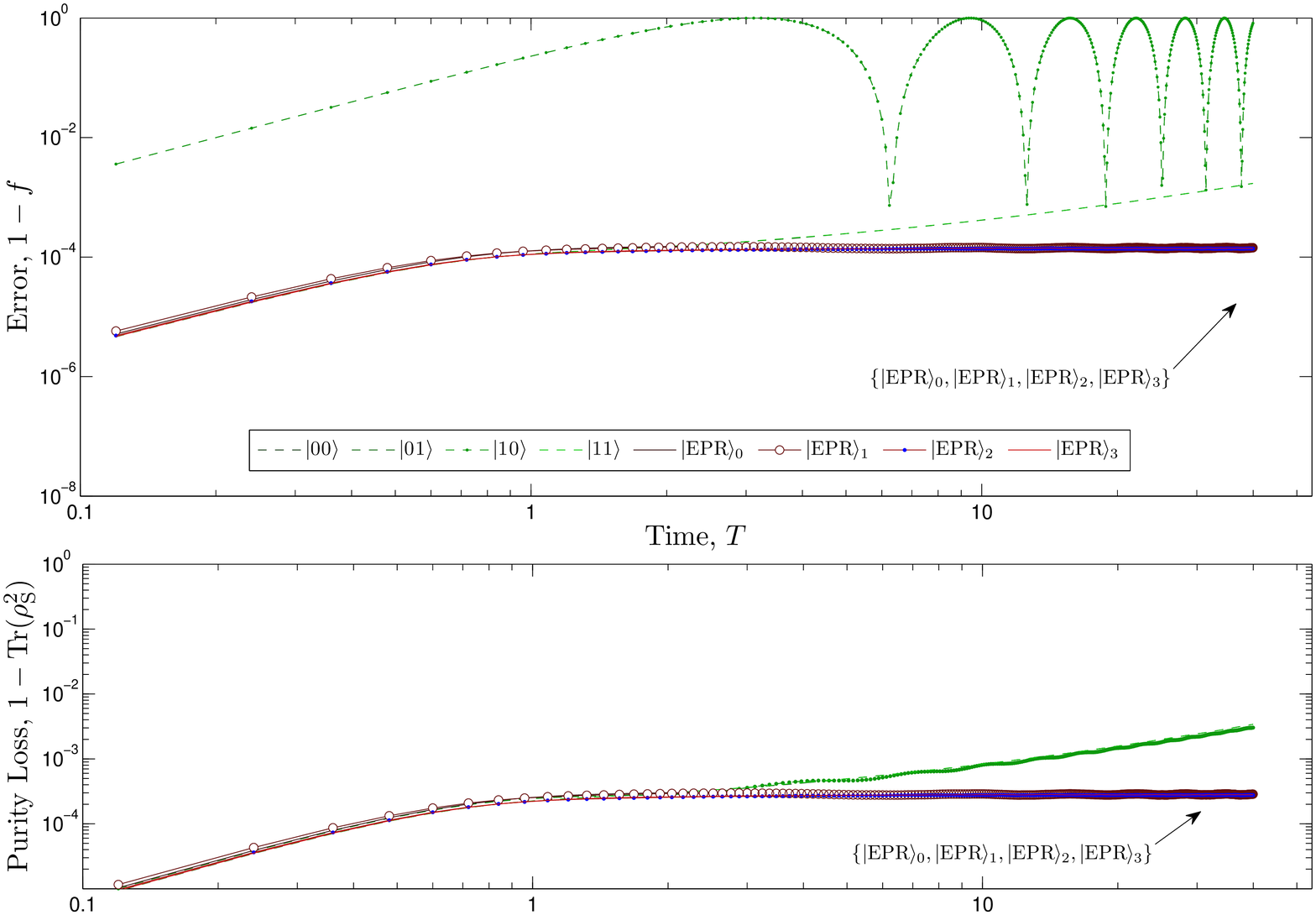} \par\end{centering}
\centering{}\caption{(Color online) Fidelity error ($1-f$) and purity
loss ($1-\text{Tr}(\rho_{S}^{2})$) for the $E3$ sequence designated to
preserve the complete Bell basis (see text). Different curves (some
overlapping) correspond to initialization in the computational basis
states and the Bell basis states, respectively. The number of bath
spins $n_{B}=15$, with $\beta =J=K$, and the pulse interval
$\tau=0.01/J$.}
\label{fig:epr-4}
\end{figure*}

\begin{figure*}[t]
\begin{centering}
\includegraphics[width=0.85\textwidth]{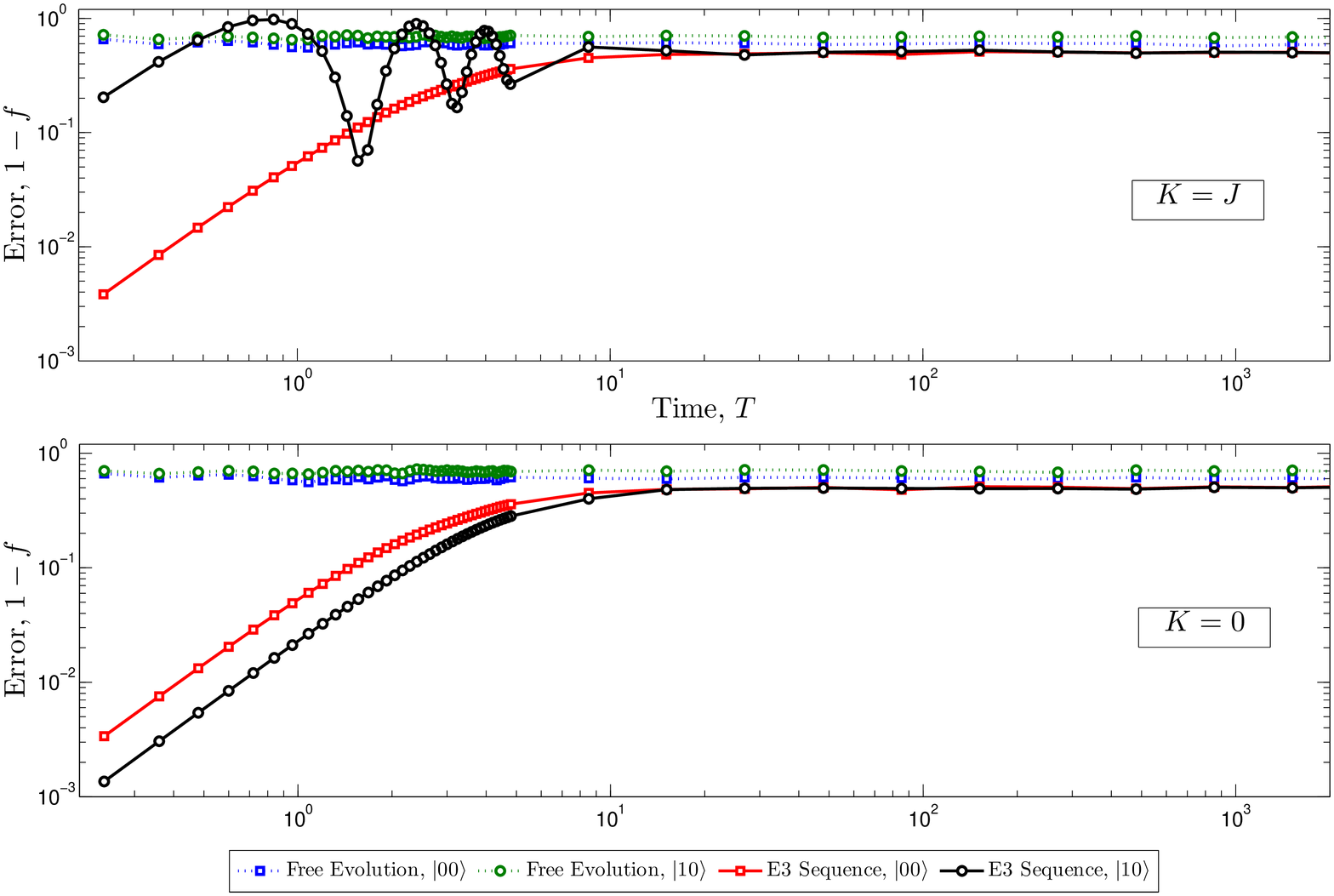} \par\end{centering}
\centering{}\vspace*{-4mm}\caption{(Color online) Comparison between
the evolution of $\ket{01}$ and $\ket{00}$ under free evolution and
the $E3$ sequence (see text) for $K=J$ (exchange-coupled qubits,
$H_S\ne0$) vs. $K=0$ (non-interacting qubits, $H_S=0$).  The number of
bath spins $n_B=8$, with $\beta =J$, and the pulse interval
$\tau=0.01/J$. Notice the oscillations of $\ket{01}$ when $K\ne0$.}
\label{fig:comp_jk}
\end{figure*}

The sequences E1, E2, and E3 were numerically simulated for the two
qubits coupled to a bath consisting of $n_{B}=15$ spins, and a set of
8 possible initializations (4 computational basis states and 4 Bell
states) was evolved for each sequence. The simulation results for E1,
E2, and E3 are summarized in Figs. \ref{fig:epr-1}, \ref{fig:epr-12},
and \ref{fig:epr-4}, respectively.  Since in this case the
contribution to $H_{SB,c}$ due to the internal exchange Hamiltonian
$H_S$ need {\em not} be fully removed by the applied control, we have
also plotted the purity loss $(1-\text{Tr}\rho_{S}^{2})$ to factor out
any remaining coherent evolution in the system and focus only on
decoherence effects when needed.  In all these figures, the eventual
differentiation of the designated PSs from the others is clearly
visible for each sequence.

Interestingly, as Fig. \ref{fig:epr-4} reveals, the E3 sequence has a
side-effect if $K \ne 0$.  We note that the intra-qubit Heisenberg
coupling $H_{S}$ has the Bell states as its eigenstates. This implies
that the E3 sequence effectively leaves $H_{S}$ invariant while it
removes the terms in $H_{SB}$ which would otherwise spoil the
eigenstates.  The net effect on the Bell states is as expected, they
are preserved as PSs. However, the Heisenberg interaction implements a
nontrivial swap-like evolution in the subspace spanned by
$\{\ket{01},\ket{10}\}$, while it acts as identity on the span of
$\{\ket{00},\ket{11}\}$.  As a result, the E3 sequence enhances the
internal system dynamics by removing the unwanted components of
$H_{SB}$, and effectively resulting in a simple ``logical gate''
\cite{Viola1999Control,Khodjasteh-Hybrid}. This explains the
oscillations in fidelity of $\{\ket{01},\ket{10}\}$ that are observed
most prominently with the E3 sequence.  These oscillations are absent
when we focus on state purity (bottom panel in Fig. \ref{fig:epr-4}).
Fig. \ref{fig:comp_jk} further highlights the difference in the
evolution of fidelity and purity by explicitly contrasting the
behavior of computational basis states in the case of interacting
vs. non-interacting qubits, qualitatively confirming the above
picture.

\subsection{Higher Order Sequences}
\label{sub:Higher-Order-Sequences}

The protocols introduced in Section \ref{sec:Control-Sequences} and
quantitatively illustrated in the above examples are all based on
cancellation of the first order (time-linear) terms in the Magnus
expansion for the effective cycle Hamiltonian $H_{c}$. Higher-order
terms in Magnus expansion can also be removed. Cancellation of
higher-order terms in the Magnus expansion is widely used in
perturbative high-order DD procedures
\cite{Khodjasteh2004,Uhrig2007,WestFongLidar2010}. The timing and
pulse type patterns used in such advanced DD schemes can be adapted to
the task of PS-engineering if desired.  While a detailed analysis is
beyond our current scope, cancellation of higher-order Magnus
corrections will result, for small enough control intervals, in
further reducing $H_{\text{per}}$ and thus the fidelity loss of the
preserved PSs.

More concretely, consider a single qubit, where instead of using the
$ZZ$ sequences of Sec. \ref{sub:Qubit-System}, we apply the reflection
$Q$ ($=\sigma_z$) follows the timing of Uhrig DD \cite{Uhrig2007}.
For higher-dimensional systems and for preserving a single PS, the
protocol described in Sec.  \ref{sub:Single-Pointer-State} can
likewise be modified to use a control cycle in which (as opposed to
using two equal length pulse intervals) the pulse intervals are
described by the Uhrig pattern \cite{Mukhtar2010}. For more than one
PS, Uhrig DD is more difficult to adapt to PS-engineering.  A possible
construction may be devised in principle for $D_S=2^{m}$-dimensional
Hilbert spaces, by adapting the results in \cite{wangprotection}.

\section{The Role of Imperfect Control}
\label{sec:imperfect}

Throughout the discussion so far, we have assumed control resources to
be {\em perfect}, allowing for precise initialization of the system in
(one of) the intended PS as well as exact implementation of all the
required control operations. We now revisit these assumptions and
analyze different ways in which limited control resources can impact
and/or modify the PS engineering problem.

\subsection{Preparation Errors}
\label{sub:Robustness-with-Respect}

A deviation of the initial state from the intended pointer basis will
propagate smoothly to the eventual fidelity of the system. Consider
for concreteness a qubit, in which the $\sigma_{z}$ eigenstates are
the designated PSs, and let the initial preparation in the PS basis be
described by 
\begin{equation}
\rho_{0}=\left(\begin{array}{cc} 1-\delta A & \delta(B+iC)\\
\delta(B-iC) & \delta A\end{array}\right),
\label{eq:rq0}
\end{equation}
where $A$, $B$, and $C$ are real numbers and the parameter $\delta$
quantifies the error strength in the preparation of $\ket{0}$.  For
simplicity, let us further assume that $\ket{0},\ket{1}$ are preserved
with the maximal fidelity 1, whereas eigenstates of $\sigma_{x}$ and
$\sigma_{y}$ evolve to a maximally mixed state as
$T\rightarrow\infty$. Using the linearity of quantum mechanics, we can
show that at $T\rightarrow\infty$, the density matrix is given by
\[
\rho_{\infty}=\left(\begin{array}{cc} 1-\delta A & 0\\ 0 & \delta
A\end{array}\right).\] 
\noindent 
The fidelity loss in the evolution of $\rho_{0}$ to $\rho_{\infty}$ up
to the leading order in $\delta$ is given by:
\begin{equation}
1-f_{\infty}=\frac{1}{2}\delta^{2}(B^{2}+C^{2})+O(\delta^{3}).
\label{eq:finf}
\end{equation}
Eq. (\ref{eq:finf}) implies that a state prepared in a sufficiently
small neighborhood of a PS (in the convex set of possible states,
including mixed ones) will evolve with a fidelity close to the maximal
fidelity. In other words, starting with a slightly misprepared initial
state and evolving under a PS-preserving protocol does result in lower
fidelities but not a complete fidelity loss.  In dynamical systems
language, the PSs (and all the mixed states diagonal in the PS basis)
are {\em Lyapunov stable} \cite{TicozziTAC,TicozziAutomatica}. The
fact that the fidelities for states prepared near PSs remains lower
than the PS-fidelities implies, however, that the dynamics is {\em not
attractive}, consistent with the purely unitary nature of the applied
control.

\subsection{Pulse Imperfections in Pointer-State Sequences}
\label{sub:Imperfect-Pointer}

While a variety of imperfections can plague control Hamiltonians in
real experiments, {\em systematic} pulse errors remain an important
limiting factor for the achievable fidelities.  As we now show, the
latter can be incorporated in our theoretical framework as long as
they result in a {\em constant} cycle propagator over the entire
control duration.  Let us reconsider a control cycle
$P_{1},\cdots,P_{n}$ designated for PS-engineering. Let $U_{P_{i}}$
denote the propagator associated with the net evolution of the open
system during (start to finish) the pulse $P_{i}$. The unitary
operators $U_{P_{i}}$ are now approximations of the {}``ideal pulses''
$P_{i}$:
\begin{equation}
U_{P_{i}}=P_{i}\exp(-iE_{P_{i}}),
\label{eq:piei}
\end{equation} 
where $E_{P_i}$ is a Hermitian system-bath operator (the so-called
``error action'' \cite{khodjasteh-2008}) that we refer to simply as
pulse error.  Following the notation of
Sec. \ref{sub:Transformation-of-Bare}, instead of Eq. (\ref{eq:ucpi})
for ideal pulses, the cycle propagator thus reads
\begin{eqnarray*} U_{c} & = &
P_{n}\exp(-iE_{P_{n}})\exp(-i\tau_{n}H)\\ & \times & \cdots
P_{1}\exp(-iE_{P_{1}})\exp(-i\tau_{1}H).
\end{eqnarray*}
The error model described by Eq. (\ref{eq:piei}) is general enough to
encompass a large class of imperfections, such as finite-width-pulse
errors (where $E_{P_{i}}$ is a system-bath Hamiltonian depending on
the pulse implementation) or, even in the narrow-pulse limit,
rotation-angle and/or rotation-axis errors (with $E_{P_{i}}$ acting on
the system only). The pulse errors $E_{P_{i}}$ are systematic in the
sense that they only depend on the pulse $P_{i}$. This results in a
fixed (constant) cycle propagator over the whole control duration,
making it still meaningful to use an effective cycle Hamiltonian
$H_{c}$ up to the total time $T$.  Letting $\varepsilon\equiv
\max_{j}\Vert E_{P_j}\Vert$ and using the Magnus expansion, we may
write [cf. Eq. (\ref{eq:magcyc})]:
\begin{eqnarray}
T_{c}H_{c} & = & \sum_{j=1}^n \tau_{j}Q_{j}^\dagger H_0 Q_{j}
+\sum_{j=1}^n Q_{j}^\dagger E_{j}Q_{j} \nonumber \\ & + &
O(T^2_{c}H_0^{2}) + O(n T_{c}H_0 \varepsilon) + O(n^{2}
\varepsilon^2), \label{eq:hcei}
\end{eqnarray}
where we have used $E_j$ for the error associated with the $j$-th
pulse in the sequence, and we have explicitly shown the dominant
linear terms while denoting the higher-order contributions by
asymptotic expressions.

The protocols for engineering PSs discussed in
Sec. \ref{sec:Control-Sequences} are all based on bringing $H_{c}$ to
obey the PS-condition of Eq. (\ref{eq:eigexp}), where
$\sum_{j}\tau_{j}Q_{j}^\dagger H_0 Q_{j}$ plays the role of
$H_{\text{dom}}$ and all the remaining terms are gathered into
$H_{\text{per}}$. Clearly, the errors associated with pulse
imperfections ($E_{P_j}$) generally belong to $H_{\text{per}}$. Thus,
while the task PS1 (ensuring PSs) can be achieved as before, the task
PS2 (ensuring higher fidelity PSs) will depend on the quality of the
pulses applied, as well as on access to shorter pulse intervals and/or
more advanced control cycles. Recall, however, that in the protocols
described in Sec. \ref{sec:Control-Sequences}, a {\em single pulse
type} ($\sigma_{p,D_S}$) is used throughout the sequence, therefore
only a single error per pulse appears in Eq. (\ref{eq:hcei}):
$E_{P_{j}} \equiv E_P$ for all $j=1,\cdots,n$. Remarkably, for such
single-pulse-type protocols, the term $\sum_{j}Q_{j}^\dagger E_{j}
Q_{j}=\sum_{j}Q_{j}^\dagger E Q_{j}$ can actually be included in
$H_{\text{dom}}$, and is {\em not} a perturbative correction. This
leaves
\begin{equation}
T_c H_{\text{per}} = T_c H^{(0)}_{\text{per}} + O( n T_{c}H_0
\varepsilon)+ O (n^2 \varepsilon^{2}),
\label{eq:same}
\end{equation}
where $H^{(0)}_{\text{per}}$ refers to ideal pulses, and thus results
in additional robustness compared to protocols that employ different
pulse types. 

\begin{figure}[t]
\begin{centering}
\includegraphics[width=0.98\columnwidth]{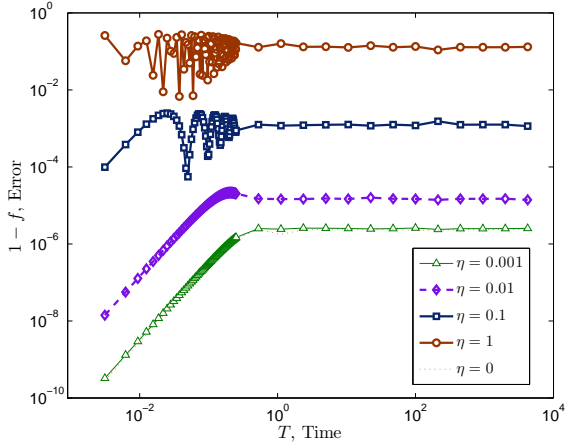}
\par\end{centering} \centering{}\caption{(Color online) Fidelity error
($1-f$) as a function of time for various over-rotation error
strengths $\eta$ for the $ZZ$ sequence used for preserving $\ket{0}$.
The error-free curve ($\eta=0$) is practically indistinguishable from
the one at low error strength ($\eta=0.001$). The couplings $\beta$
are in units of $J$, and $\beta=J$. The pulse interval is fixed at
$\tau=0.01/J$ and the number of bath spins $n_{B}=7$.}
\label{fig:sys_err}
\end{figure}

The crucial implication of the redefinition of $H_{\text{dom}}$
vs. $H_{\text{per}}$ in Eqs. (\ref{eq:hcei}) and (\ref{eq:same}) is
that although this will generally imply a worse fidelity lower-bound
{[}cf. Eq. (\ref{eq:1mfguar1}){]} for a PS-protocol implemented with
imperfect pulses, such imperfections will {\em not} result in a
degradation of fidelity with time.  This simple yet practically
important point is quantitatively illustrated in
Fig. \ref{fig:sys_err} for a single qubit interacting with a spin bath
as in Sec. \ref{sub:Qubit-System}.  Specifically, we have analyzed the
combined effect of rotation-angle and rotation-axis errors in
preserving the state $\ket{0}$ the $ZZ$ protocol: the pulse error is
characterized by letting $E_P \equiv
\eta(e_{x}\sigma_{x}+e_{y}\sigma_{y}+e_{z}\sigma_{z})$, where $\eta\in
[0,\pi]$ characterizes the over-rotation strength, and
$e_{x},e_{y},e_{z}$ are arbitrary (but fixed for all the data points)
numbers randomly sampled from $[-1,1]$ which characterize the axis
misalignment.  The observed {\em robustness of the engineered PSs}
with respect to systematic pulse errors is in sharp contrast to the
behavior of DD protocols where repeating control cycles with faulty
control tends to constantly degrade state fidelities for an
arbitrarily chosen state \cite{Tyryshkin2010}. 

\section{Pointer States from Imperfect Decoupling Sequences}
\label{sub:slava}

In Sec. \ref{sub:Qubit-System}, we showed that the {\em imperfect
cancellation} associated with single-qubit universal DD sequences
results in the accidental generation of PSs.  For instance, the $XYXY$
sequence leads to a pointer basis in which the $Z$ eigenstates are
preserved and, as shown in the previous section, can survive without
significant fidelity loss even in the presence of systematic control
errors.  Interestingly, however, even in situations where decoupling
is theoretically exact, PSs can still emerge from \emph{imperfect
pulses}. This manifests in a strongly state-dependent degree of
stability after many DD cycles, as recently demonstrated in the
context of DD experiments using the electron spin resonance (ESR) of
phosphorus donor spins in silicon, see in particular the data in
Fig. 2 of \cite{Tyryshkin2010} and Fig. 1 of \cite{Wang2010}.  In this
Section, we show how the emergence of these stable states can be
naturally re-interpreted and analyzed within our general framework, in
terms of the emergence of a PS-supporting effective Hamiltonian.

The system under consideration is described in detail elsewhere
\cite{Tyryshkin2010,Wang2010}.  For a given spin $S$, in a frame that
rotates with the electron's Larmor frequency, the bare Hamiltonian
describing the interaction between each $P$ electron spin and the
nuclear spin bath can be approximated semiclassically approximated by
[See Sec. \ref{sub:Classical-Environment}]
\[ H_{SB}=\gamma_e S_z b_{z}, \] 
\noindent 
where $\gamma_e$ and $b_z$ are the electronic gyromagnetic ratio and
the total effective magnetic field, respectively.  The latter accounts
for the effect of the nuclear spin environment and the off-resonance
error resulting from spatial inhomogeneity across the sample, both of
which can be treated as a random variable static in time to a good
approximation.  Clearly, the above interaction Hamiltonian implies a
``natural'' pointer basis along the $Z$ axis in the absence of
control: besides the eigenstates of $\sigma_{z}$, other states will
not be preserved.  
In the presence of perfect (instantaneous) pulses, a universal DD
cycle (such as $XYXY$ or $XZXZ$) removes $H_{SB}$ exactly. However, in
the actual experiments, the DD pulses are not ideal, and their
imperfections can rapidly accumulate over time, eventually nullifying
the expected preserving action of DD on the spin's quantum
state. Despite their destructive role in DD, pulse imperfections can
in fact select PSs, provided that they are constant over time,
although inhomogeneous across the sample \cite{slava}. For ESR
experiments, where the pulses are indeed close to ideal, we can
approximate them as unitary rotations, with
the rotation axes and angles slightly differing from their nominal
values. In particular, for $X$ and $Y$ pulses, following the notation
of Refs. \cite{Tyryshkin2010,Wang2010} the actual evolution operators
describing the imperfect rotations are
\begin{eqnarray*}
\label{eq:rotationXY}
U_X&=&\exp{[-i(\pi + \epsilon_x)(\frac{1}{2}\vec \sigma\cdot\vec n)]},
\\ \nonumber U_Y&=&\exp{[-i(\pi + \epsilon_y)(\frac{1}{2}\vec
\sigma\cdot\vec m)]},
\end{eqnarray*}
where $\epsilon_{x(y)}$, with $ \epsilon_{x} \approx \epsilon_{x}
\equiv \epsilon$, are small errors in the rotation angles, and $\vec
n, \vec m$ are the actual rotation axes slightly differing from their
nominal directions along $x$ and $y$, respectively, that is, we may
write $\vec n \equiv (\sqrt{1-\eta^2 n^2_y -\eta^2 n^2_z},\eta
n_y,\eta n_z)$ and $\vec m \equiv (\eta m_x,\sqrt{1-\eta^2
m^2_x-\eta^2 m^2_z},m_z)$, with $\eta \ll 1$.  The parameters
$\epsilon$ and $\eta$ thus play the role of the perturbative parameter
$\epsilon$ in the previous sections. The simplified but realistic
one-dimensional model derived for ESR on a Si:P sample
\cite{Tyryshkin2010,Wang2010} suggests that the probability
distribution for the rotation angle error is peaked around a value
$\epsilon_0 \geq 0$ and is given by
$P(\epsilon)=(1/2\epsilon_0)[3(1-\epsilon/\epsilon_0)]^{-1/2}$, with
$-2\epsilon_0 \leq \epsilon \leq \epsilon_0$.  The axis offsets $\eta
n_{y (z)}$ and $\eta m_{x(z)}$ each follow a similar distribution
peaked around $n_0 \geq 0$.

Using Eq.~(\ref{eq:ucyc}), we can formally define the effective cycle
Hamiltonian $H_c$ associated with applying the $XYXY$ sequence at the
inter-pulse interval $\tau$ (with $T_c=4\tau$). By direct calculation,
taking into account terms up to the second order in pulse errors, we
obtain:
\begin{eqnarray*}
T_c H_c &\hspace*{-2mm}=\hspace*{-1mm}& \sigma_z [ -2 \eta
(m_x+n_y)+(\epsilon^2/2) \cos{b_z\tau}] \\ & \hspace*{-2mm}-
\hspace*{-1mm} & 2 \eta (m_x+n_y) \{ [ (\epsilon/2) (1 +\sin{b_z\tau})
  - \eta n_z \cos{b_z\tau} ] \sigma_x \\ & \hspace*{-2mm}
+\hspace*{-1mm} & [\eta m_z - (\epsilon/2) \cos{b_z\tau} - \eta
  n_z\sin{b_z\tau} ]\sigma_y \}.
\end{eqnarray*}
From the above expression, one can see how the separation of $H_c$
into the dominant and the perturbative term happens.  As long as
off-axis errors are present ($\eta \ne 0$, independently of whether
rotation errors are present also), the dominant term is simply the
first contribution, $T_c H_{\rm dom}= -2 \eta
\left(m_x+n_y\right)\sigma_z$, with all remaining terms belonging to
the perturbative Hamiltonian $H_\text{per}$, which is $O[\eta
\max(\epsilon, \eta)]$ and is not in any particular direction.  Using
the results of Sec. \ref{sub:Semi-Classical-Bath}, we expect the
fidelity loss of PSs ($Z$ eigenstates) to be proportional to the
square of the ratio of the perturbative to dominant Hamiltonian, and
hence controlled by $O[\max(\epsilon, \eta)]$.  In the case where
$\eta=0$, to the leading order in the perturbative parameters we have
instead $T_c H_c = (\epsilon^2/2) (\cos{b_z\tau})\sigma_z $. Thus,
while it is formally necessary to redefine $H_{\rm dom}$ so that it
includes second-order contributions, the generated PSs will still be
the $Z$ eigenstates.

The $XZXZ$ protocol is equivalent to the $XYXY$ one in the absence of
pulse errors but when pulse errors are taken into account, the
resulting effective cycle Hamiltonian becomes qualitatively different:
\begin{eqnarray*}
T_c H_c &\hspace*{-2mm}=\hspace*{-1mm} & \sigma_{y} [ (\epsilon/2) (1-
\sin{b_z\tau}) - \eta n_z (1-\cos{b_z\tau}) ] \\
&\hspace*{-2mm}-\hspace*{-1mm} & [(\epsilon/2) (1- \sin{b_z\tau}) -
\eta n_z (1-\cos{b_z\tau}) ] \\ \nonumber &\hspace*{-2mm}\times
\hspace*{-1mm} &\{ -\eta m_x\sigma_x + [(\epsilon/2)(1+\cos{b_z\tau})
  - \eta m_z \\ \nonumber &\hspace*{-2mm}+\hspace*{-1mm}& \eta
  n_z\sin{b_z\tau}] \sigma_z \},
\end{eqnarray*}
where we took into account that $Z$ pulses in ESR experiments are
implemented as $Z=X\,Y$, by using two closely spaced $X$ and $Y$
pulses \cite{Tyryshkin2010}.  Clearly, the dominant term is the one in
the first row of the above equation, which includes only terms of the
first order in the pulse errors, and implies that the $Y$ eigenstates
now emerge as the pointer basis.  The perturbative correction is
$O[\max(\epsilon, \eta)^2]$ and is not in any particular direction.
Again, the fidelity loss is proportional to the ratio of the
perturbative to dominant Hamiltonian, and controlled by
$O[\max(\eta,\delta)]$.

This analysis parallels the consideration of
Sec.~\ref{sub:Qubit-System}.  Thus, it is not surprising that the
qualitative conclusions about the character of the generated PS are in
perfect agreement with the experimental and theoretical results
reported in \cite{Tyryshkin2010,Wang2010}.  A quantititative analysis
would need to take into account additional factors, most importantly
the need to average the fidelity bound over the probability
distributions of the offsets and over-rotation errors [similar to
Eq. (\ref{eq:fNhxhyhz})].  While additional technical analysis of
these specific experiments goes beyond our current scope, the above
clearly demonstrates how our general framework can be modified to
include the PS generated by DD pulse imperfections in context of
direct experimental relevance.

\section{Conclusion}
\label{sec:Conclusion}

We have provided general open-loop unitary control protocols for
engineering the interaction between a finite-dimensional system and
its environment in the non-Markovian limit, in such a way that a
desired set of pure states can be maintained as effective pointer
states (PSs) of the dynamics.  Our constructive results are supported
by analytical upper bounds for the fidelity loss of the engineered
PSs, which allow for systematic improvement by simply employing faster
and/or more elaborated control schemes.  While similar in flavor to
dynamical decoupling protocols for protecting arbitrary quantum
superpositions, the methods presented and analyzed in this paper aim
to provide {\em selective energetic protection} for a designated set
of pure states and their {\em convex} combinations only, effectively
allowing for the on-demand generation of a {\em robust classical
memory}.  The fact that PSs are protected via relative energy gaps in
the interaction with the environment is manifest in the nature of our
performance bounds, which rely on von Neumann mean ergodic theorem and
simple perturbation theory.  Physically, the key requirement is to
synthesize an effective Hamiltonian with a desired ``dominant''
symmetry structure, and to reduce the ``perturbing interactions'' that
would otherwise destabilize the state and thus reduce the fidelity.

At the expenses of making the control design state-dependent, the
PS-engineering protocols we have introduced have distinctive
advantages over general-purpose dynamical decoupling schemes, which
become most transparent in the task of engineering complex quantum
states in multi-qubit systems as PSs.  While the task of Bell-state
engineering we have analyzed in depth provides a paradigmatic example
in this respect, the ability to engineer arbitrary PSs can potentially
be useful as a resource in quantum information processing and/or
quantum metrology tasks.  It is also worth stressing that the methods
we have presented are directly accessible via pulsed control, but can
modified in principle to allow continuous-time protocols that still
effectively transform the interaction with the environment.
Ultimately, stronger controls or faster pulse rates are the basic
resources we leverage in achieving high-fidelity PSs within our
Hamiltonian open quantum system setting.

A number of further questions may be worth addressing.  For instance,
it may be interesting to examine whether the present framework can be
extended to encompass the more general symmetry conditions that allow
for {\em time-dependent} PSs, as recently considered in \cite{Drake}.
From a control-standpoint, the latter problem might in turn relate to
the possibility of robust time-dependent state-tracking, rather than
long-time state-preservation as examined here.  Interestingly, the
distance of a density operator has recently be invoked to quantify
``quantumness'' in the context of coherent energy transfer in
biological systems \cite{Nalbach}.  It is both natural and intriguing
to ask whether a perturbative mechanism similar to the one involved in
the generation of stable PSs examined here may be brought to bear on
the problem of further understanding long-lived quantum coherences in
complex systems.
Lastly, the energetic protection against decoherence enjoyed by the
PSs may be ultimately connected to quantum noiseless subsystem codes.
Thus, a natural (although possibly highly non-trivial) direction for
exploration is whether the idea of an engineered fidelity guarantee
can be viable for a subsystem code that can preserve genuine quantum
information, as opposed to only classical information as for a PS
basis.

\begin{acknowledgments}
L.V. gratefully acknowledges support from the NSF through award
No. PHY-0903727.  It is a pleasure to thank Francesco Ticozzi and
Winton G. Brown for insightful discussions during the course of this
work.  Work at Ames Laboratory was supported by the Department of
Energy --- Basic Energy Sciences under Contract No.~DE-AC02-07CH11358.
\end{acknowledgments}

\appendix

\section{Mean Ergodic Theorem}
\label{sec:Mean-Ergodic-Theorem}

Let $U$ and $X$ denote, respectively, a unitary operator and arbitrary
bounded operator acting on the same Hilbert space. The {\em von
Neumann's Mean Ergodic Theorem} (MET) implies that the limit
\[
\lim_{N\rightarrow\infty}\frac{1}{N}\sum_{n=0}^{N-1}U^{-n}X U^{n}
\equiv X^{\Vert}\]
\noindent 
exists and satisfies $[X^{\Vert},U]=0$ \cite{reed1980functional}.  Note that convergence is
implied in any inner product distance between operators that is
invariant under the adjoint action $X\mapsto U^{\dagger}XU$.  In the
above form, the MET has also been invoked to formally relate DD to the
quantum Zeno effect \cite{Facchi2004}.

To visualize the theorem for a finite-dimensional system, consider the
complete basis of eigenstates $\{\ket{\phi_{i}}\}$ of $U$ with
eigenvalues $e^{i\phi_{i}}$. The operator $X$ can be written in the
basis of the $E_{ij}=\ket{\phi_{i}}\!\bra{\phi_{j}}$ operators. The
latter are transformed according to
\[ U^{-n}E_{ij}U^{n}\mapsto e^{in(\phi_{i}-\phi_{j})}E_{ij} .\] 
\noindent 
Consequently, as long as there is no degeneracy ($\phi_{i}\ne\phi_{j}$
unless $i=j$), we have
\begin{equation}
\frac{1}{N}\sum_{n=0}^{N-1}U^{-n}E_{ij}U^{n}=\frac{1}{N}
\frac{ 1-e^{ i(N+1)(\phi_{i}-\phi_{j}) } }{1-e^{i(\phi_{i}-\phi_{j})}} 
E_{ij}, 
\label{eq:geomsum}
\end{equation}
where in the limit of $N\rightarrow\infty$, the r.h.s. approaches zero
when $i\ne j$ and 1 otherwise. Thus, only diagonal basis elements are
preserved under averaging and the remaining ones are
annihilated. These diagonal basis elements span operators that commute
with $U$. Notice that the left-over terms with $i\ne j$ in the limit
of a shrinking minimum gap, $\Omega=\min_{i\ne
j}\vert\phi_{i}-\phi_{j}\vert$, scale with $1/(N\Omega).$

\section{Initial Fidelity Decay}
\label{sec:Appendix:Initial-Decay}
While the goal of our scheme is long-time manipulation of coherence
using periodically repeated control cycles, we can nonetheless
approximate the initial short time behavior of the system. In what
follows we assume that
\[ N\delta \equiv N  T_{c}\epsilon \Vert
H_{\text{per}}\Vert\ll 1 ,\]
\noindent 
where $N$ is the number of control cycles that have been applied to
the system up to time $T=NT_c$. This allows to use the linear (in
$\delta$) component of the interaction picture propagator
$\exp(-i\Omega^{[1]}_N)$ as a substitute for the propagator when PS
fidelity is concerned. We can estimate the norm of the effective
Hamiltonian $\Omega^{[1]}_N$ even before using Eq. (\ref{eq:geomsum}),
directly from Eq. (\ref{eq:omega1}) in the main text:
\[
\Vert \Omega^{[1]}_N\Vert \le N\Vert E\Vert \approx N\Vert
H_\text{per}\Vert=N\delta.  \]

The fidelity loss of the system after $N$ cycles can be bounded
starting from the following bound for the Uhlmann fidelity $f_N^U
=f_N^{1/2}$\cite{fuchs99}:
\[
 1-f_{N}^U \leq D\Big[\ket{0}\!\bra{0},\exp(-i\Omega^{[1]}_N)\ket{0}\!
\bra{0}\exp(+i\Omega^{[1]}_N)\Big] , \]
\noindent 
where $D[\rho_{1},\rho_{2}]$ is the trace distance:
\[ D[\rho_{1},\rho_{2}]\equiv \frac{1}{2}\Vert\rho_{1}-\rho_{2}\Vert_1 .\]
\noindent 
We can then use $(1-f_N)\le 2(1-f_{N}^U)$ [following from $1-x\le
2(1-x^{1/2})$ for $0<x<1$] and the general bounds in
Ref. \cite{Lidar-Bounds}, to show that
\begin{align}
1-f_N & \le \exp(\Vert \Omega^{[1]}_N+O(N^{2}\delta^{2})\Vert)-1
\nonumber \\ & \le (e-1)\Vert \Omega^{[1]}_N \Vert \le (e-1)N \delta ,
\label{Bbound}
\end{align} 
where in the last step we used $e^{x}-1<(e-1)x$, for $x\le1$.  We have
thus bounded the short-time fidelity loss in terms of a linear
function of the number of cycles $N$ for small $N$. Note that for a
single cycle, a tighter fidelity bound can be given, directly in terms
of the (bath-averaged) variance of $H_{\text{per}}$ in the initial PS
\cite{Wen08}, leading to $1-f(T_c) \lesssim (\epsilon \Vert
H_{\text{per}} \Vert)^2 T_c^2$.  For large values of $N$, the results
of Sec. \ref{sec:Ergodic-Averaging-Induced} are applicable instead.

\section{Implementing Arbitrary Reflections}
\label{sec:appendix:reflect}
Implementing the PS-preserving pulses defined in
Sec. \ref{sec:Control-Sequences} is straightforward provided that
tunable control Hamiltonians diagonal in the pointer basis are
available, that is, Hamiltonians of the form
\[ H_{\text{ctrl}}(t)=\sum_{i=0}^{D_{S}-1}h_{i}(t)\ket{i}\!\bra{i},\]
Note that in principle, even a constant in time Hamiltonian which is
diagonal in the pointer basis and has non-commensurate eigenvalues
would suffice for generating any unitary which is diagonal in the
pointer basis \cite{NielsenBook}.  If, however, no such control
Hamiltonians are available, the problem of producing the required
pulse operators can be difficult.

In particular, the pulse operators used in our protocols can be
entangling if the system is multipartite. For example, this was
explicitly the case in the PS-protocols employed for Bell-state
engineering, Sec. \ref{sub:EPR-Basis-Preservation}.  Furthermore, note
that the operator $Q$ used in Sec. \ref{sub:Single-Pointer-State} to
preserve a single PS corresponds to a multiple-controlled-phase
quantum gate, thus, generally, entangling operations are required also
to preserve a single product state (say, an element of the
computational basis).  Interestingly, however, preserving all
computational basis elements in a multiple-qubit system requires no
entanglement.  For example, consider the control cycle resulting from
concatenating (nesting) the $ZZ$ cycles acting on every qubit
\cite{Khodjasteh2004,wangprotection}, resulting in preservation of the
computational basis as an engineered pointer basis. These ideas
suggest possible connections to the quantum search algorithm
\cite{Grover:1996:FQM:237814.237866}, in which the oracle performs the
same (entangling) operation $Q$ to select a particular element in the
computational basis.

\bibliographystyle{apsrev}
\bibliography{qip}

\begin{thebibliography}{77}
\bibitem[{sid()}]{sidenote}
\bibinfo{note}{The assumption in Eq. (\ref{residual}) appears to be used for
  simplifying the algebra but it hinges upon the knowledge of the dominant
  Hamiltonian and its PS set. This will be crucial to examine the unintended
  PSs emerging in DD protocols, see Sec. \ref{sub:Qubit-System}.}
\bibitem[{sla()}]{slava}
\bibinfo{note}{Note that each spin undergoes a slightly different evolution
  when subjected to an imperfect DD pulse. Strictly speaking, in such a
  situation, a PS should be understood slightly differently. It is no longer a
  pure state describing a single quantum spin, but rather the density matrix
  describing an ensemble of identical spins. The component of this density
  matrix which remains stable under the action of the pulses has a diagonal
  form in the basis of stable states. Therefore, a set of PS $|j\rangle$ can
  still be formally associated with it, although these PS do not describe an
  actual quantum state of a single spin. Alternatively, one may consider a
  quantum state of an ensemble of independent spins, and define a set of
  product PSs $|J\rangle = \otimes_{k=1}^n |j\rangle_k$, where $k=1,\dots n$
  indexes the spins belonging to an ensemble. In this case, we are dealing with
  $2^n$ PSs of the spin ensemble.}
\end{thebibliography}

\end{document}